\documentclass[12pt,onecolumn,accepted=2017-12-11]{quantumarticle}
\usepackage[numbers,sort&compress]{natbib}
\usepackage[colorlinks]{hyperref}
\usepackage{amsmath, mathtools, amsfonts, dsfont}

\title{BQP-completeness of Scattering in Scalar Quantum Field Theory}

\author[1,2]{Stephen P. Jordan}
\author[3]{Hari Krovi}
\author[4]{Keith S. M. Lee}
\author[5]{John Preskill}
\affil[1]{\small{National Institute of Standards and Technology,
               Gaithersburg, MD, USA}}
\affil[2]{\small{Joint Center for Quantum Information and Computer Science, University of Maryland, College Park, MD, USA}}
\affil[3]{\small{Quantum Information Processing Group, Raytheon BBN Technologies, Cambridge, MA, USA}}
\affil[4]{\small{Centre for Quantum Information \& Quantum Control and
Department of Physics, University of Toronto, Toronto, ON, Canada}}
\affil[5]{\small{Institute for Quantum Information and Matter,
California Institute of Technology, Pasadena, CA, USA}}

\begin{document}

\bibliographystyle{plainnat}

\newcommand{\bra}[1]{\langle #1|}
\newcommand{\ket}[1]{|#1\rangle}
\newcommand{\eq}[1]{Eq.~(\ref{#1})}
\newcommand{\sumint}[1]{\quad \mathclap{\displaystyle\int}\mathclap{\textstyle\sum_{#1}}\;\;\;}
\renewcommand{\th}{^\mathrm{th}}
\newcommand{\id}{\mathds{1}}
\newcommand{\sect}[1]{\S\ref{#1}}

\maketitle

\begin{abstract}
Recent work has shown that quantum computers can compute scattering 
probabilities in massive quantum field theories, with a run time that is
polynomial in the number of particles, their energy, and the desired 
precision. Here we study a closely related quantum field-theoretical problem: 
estimating the vacuum-to-vacuum transition amplitude, in the presence of 
spacetime-dependent classical sources, for a massive scalar field theory in 
$(1{+}1)$ dimensions.  We show that this problem is BQP-hard; in other words, 
its solution enables one to solve any problem that is solvable in polynomial 
time by a quantum computer. Hence, the vacuum-to-vacuum amplitude cannot be 
accurately estimated by any efficient classical algorithm, even if the field 
theory is very weakly coupled, unless BQP=BPP. Furthermore, the corresponding 
decision problem can be solved by a quantum computer in a time scaling 
polynomially with the number of bits needed to specify the classical source 
fields, and this problem is therefore BQP-complete. 
Our construction can be regarded as an idealized architecture for a universal 
quantum computer in a laboratory system described by massive $\phi^4$ theory  
coupled to classical spacetime-dependent sources.
\end{abstract}

\section{Introduction}

The field of computational complexity theory is the study of the 
resources required to solve computational problems.  
Problems with the same intrinsic difficulty are categorized into complexity 
classes, which can be either classical or quantum,
and relationships between different classes are studied.
The class of computational problems that are solvable in polynomial 
time by quantum computers, with a small probability of error, is 
called BQP. The class of problems solvable in polynomial time by classical
randomized computations with a small probability of error is called BPP. It is
conjectured that BPP is equal to P, the class of problems solvable
with certainty by deterministic classical computers
(see, for example, \cite{Arora-Barak}). 

A BQP-hard problem $\mathcal{P}$ is one with the property that any problem
in BQP can be efficiently translated into an instance of
$\mathcal{P}$, so that the answer to the instance of $\mathcal{P}$
gives the answer to the original problem. The method for efficient
translation is required to be a polynomial-time classical computation
and is referred to as a \emph{reduction}. A simple example of a
BQP-hard problem is the following: given a bit string describing a quantum 
circuit $C$, decide whether the corresponding unitary operator $U_C$ has an 
all-zeros to all-zeros transition probability 
$|\bra{0\ldots0} U_C \ket{0 \ldots 0}|^2$ greater than $2/3$ or smaller than
$1/3$, if one of these is guaranteed to be the case. We use a
reduction from this problem to a problem of estimating
vacuum-to-vacuum transition probabilities in a quantum field theory to
show that the latter is also BQP-hard. If any BQP-hard problem were 
solvable in polynomial time by classical computers, then all of
quantum computation would be efficiently simulable classically. (Thus,
BPP would equal BQP.) It is widely believed that this is impossible
and, therefore, if a problem is BQP-hard it is intractable for
classical computers.

We show BQP-hardness for the problem of computing a vacuum-to-vacuum
transition probability in $(1+1)$-dimensional $\phi^4$ theory with 
spacetime-dependent external fields. Specifically, suppose that initially
the quantum field theory is in its vacuum state and all external
fields are turned off. Then, the external fields are applied with some
specified variation in spacetime. Eventually, the external fields are
again turned off. The computational problem is to calculate whether
the final state of the system is the vacuum. More precisely, the system, 
being quantum mechanical, can be in a superposition of the vacuum state and
other states, and the problem is to decide whether the probability
(that is, squared amplitude) of being in the vacuum state is large or
small.

In previous work \cite{JLP12, JLP14, JLP14b}, we showed that quantum
computers can efficiently compute transition probabilities in 
certain interacting quantum field theories, including $\phi^4$ theory. 
Here, we show that a slight variant of the problem solved in \cite{JLP12,JLP14} 
is BQP-hard. Essentially, this result implies that classical computers
cannot solve the problem in polynomial time unless BQP=BPP, and thus
the quantum algorithm of \cite{JLP12,JLP14} constitutes a genuine
superpolynomial speedup. The scattering process used in our BQP-hardness
construction differs from the process simulated in
\cite{JLP12,JLP14} in that spacetime-dependent source terms are present. 
Nevertheless, by standard arguments \cite{suzuki:1993, wiebe:2010}, such 
terms at worst induce modest efficiency penalties
on the Suzuki-Trotter formulae used in \cite{JLP12,JLP14}. A second
difference is that the BQP-hard problem introduced here is to estimate
a vacuum-to-vacuum transition probability, whereas the simulation
algorithm of \cite{JLP12,JLP14} samples from a probability distribution
defined by a set of local measurements. 
From the methods introduced in
\cite{JLP12,JLP14} for implementing the unitary time evolution and preparing
the vacuum state with efficient quantum circuits, one can construct an
efficient quantum algorithm estimating the vacuum-to-vacuum transition
probability using the \emph{Hadamard test}.  
Thus, the algorithm of \cite{JLP12,JLP14} 
suffices to show that the BQP-hard transition-probability decision
problem discussed here is also contained in BQP. Problems such as
this, which are both BQP-hard and contained in BQP, are called
BQP-complete.


The quantum field theory we consider is described by the Lagrangian
\begin{equation}
\label{lagrangian}
\mathcal{L} = \frac{1}{2} \partial_\mu \phi \partial^\mu \phi
           - \frac{1}{2} m^2 \phi^2 - \frac{1}{4 !}\lambda \phi^4
           - J_2 \phi^2 - J_1 \phi 
\,,
\end{equation}
where $J_1 = J_1(t,x)$ and $J_2= J_2(t,x)$ are the external fields. We consider
the computational problem of, given bit strings\footnote{The functions
  $J_1(t,x)$ and $J_2(t,x)$ have bounded spatial extent and limited
  bandwidth, and therefore they can be specified with polynomially
  many bits (see Sec.~\ref{sec:overview}).}
specifying $J_1(t,x)$ and $J_2(t,x)$, predicting whether the system remains 
in the vacuum state. Specifically, at time zero the sources $J_2$ and $J_1$ 
are zero and the system is in the vacuum state. Then $J_2$ and $J_1$ are 
varied in time as specified by the given bit strings and return to zero at 
time $T$. The computational problem is to decide whether the probability of 
remaining in the vacuum state at time $T$ is greater than $2/3$ or smaller 
than $1/3$, given a promise that one of these is the case. The constants $1/3$ 
and $2/3$ are conventional but arbitrary; our hardness result 
would be unchanged for other choices.

From the perspective of scientific computing, this formulation of the 
problem perhaps seems unusual. 
In real applications, one typically wants to compute a quantity of interest
to within some precision $\epsilon$, a task referred to here
as an estimation problem. 
However, decision problems (namely, those whose answers are either 
``yes'' or ``no'') 
are more convenient for complexity theory, and hardness results for
decision problems automatically imply the hardness of corresponding, 
more natural, estimation problems. Clearly, if one could solve the estimation
problem of computing the vacuum-to-vacuum transition probability $p$
to within $\pm \epsilon$ for some $\epsilon < 1/6$, then one could use
this to answer the decision problem of whether $p < 1/3$ or
$p > 2/3$. Thus, our BQP-hardness result implies that neither of these
problems can be solved in polynomial time by classical computers,
as long as BPP $\neq$ BQP.


Previous work has investigated the computational complexity of
approximating scattering amplitudes for particles hopping among the
vertices of a graph \cite{CGW13, BHST14, C09}. The techniques
developed in these earlier works could be relevant to BQP-hardness
constructions for quantum field theories, especially if one is 
interested in external fields that are time-independent. However, 
in quantum field-theoretical scattering, one is faced with problems
not encountered in scattering on graphs, in particular the
encoding of the problem instance. 
In graph scattering, the instance is typically encoded in the graph. 
Here, we encode it in the spacetime dependence of an external field. 
Also, the graph serves to confine the particles.

The rest of this paper is organized as follows.
Section~\ref{sec:overview} presents an overview of our construction
and a discussion of our results.
In Sec.~\ref{sec:prep}, we show how a state representing an 
initialized array of qubits can be prepared with arbitrarily high
fidelity.
Then, in Sec.~\ref{sec:gates}, we describe how to implement a
universal set of quantum gates. The two-qubit gates we construct are subject to leakage from the computational subspace, and we explain how this issue can be addressed. 
Finally, in Sec.~\ref{sec:measurement}, we discuss the Hadamard test
and particle-detector measurements, two different means of obtaining 
BQP-completeness results. 
Some technical details are relegated to appendices.

\section{Overview and Discussion}
\label{sec:overview}

\subsection{Choice of problem}
There are potentially many different computational problems arising in
quantum field theory whose hardness one might wish to study. 
In choosing a problem to establish the BQP-hardness of, we have been
guided by the criterion that the problem should be physically natural.
In other words, it should be as close as possible to familiar problems 
one is interested in solving in practice. 
Specifically, the choice entails the selection of a particular field theory
and the set of allowed inputs and observables.
These must have sufficient richness to allow the encoding of a quantum 
circuit whose output is to be ``simulated'' by the dynamics of the quantum 
field theory. 
From the computational perspective, of course, the more economical the 
choice is, the stronger and more interesting an associated BQP-hardness
result will be.
With these factors in mind, we consider the problem of
computing a vacuum-to-vacuum transition amplitude in a theory with
spacetime-dependent external fields, where the description of the 
external fields constitutes the input to the problem. Such a calculation 
is the evaluation of a generating functional $Z[J]$. 
The formal computational
problem that we have proposed and analyzed in this paper is physically
and computationally well-motivated. However, other reasonable
BQP-hardness statements can be proposed, not all of which are
manifestly equivalent to ours.

In particular, we have defined scattering to be purely unitary dynamics 
without any measurements performed during the scattering process. This is 
in keeping with the standard notion of scattering in quantum field theory. 
If intermediate measurements and feedforward are allowed, then simpler 
BQP-hardness constructions may be possible along the lines of the KLM 
construction \cite{KLM01}. In architectures for real quantum computers, 
intermediate measurements and active error correction are used to achieve 
a constant success probability in quantum computations of polynomially many 
gates, even though each gate is implemented with only constant precision. 
In our BQP-hardness construction, we instead achieve a constant success 
probability by implementing each of the $G$ quantum gates with an infidelity 
scaling as $O(1/G)$ and preparing each of the $n$ qubits with an infidelity 
scaling as $O(1/n)$.

Our definition of the scattering problem allows spacetime-dependent
source terms, which break translational invariance, in the Lagrangian
of the quantum field theory. Physically, such source terms correspond 
to externally applied classical fields. In other words, although
the laws of physics are invariant under translations in time and
space, the presence of an experimental apparatus in a particular location
breaks this symmetry. Our formulation of the scattering problem 
considers the experimental apparatus that applies the fields
that manipulate the qubits to be external. We do not demand quantum 
field-theoretical simulation of the particles making up this apparatus. 

Lastly, in our BQP-hardness construction, we have demanded that the
initial state be the vacuum. The creation of the particles to be scattered
is considered part of the dynamics. This makes our construction more
complicated, as we must design a state-preparation scheme and analyze
its fidelity (\S\ref{sec:prep}). Our construction implies as an
immediate corollary that, if one allows the initial state in the
scattering problem to consist of particles bound in the potential
wells, then the associated scattering problem is BQP-hard. By showing 
that BQP-hardness still holds when the initial state is restricted to be 
the vacuum, we achieve a meaningful strengthening of our result. 
In a high-energy scattering problem, one is typically interested in
situations where there are initial-state particles, but these are 
propagating relativistically, rather than already bound in potential wells.  

One can heuristically obtain a BQP-hardness result for the Standard Model 
of particle physics by noting that the physics accessible in today's 
laboratories is described by the Standard Model. 
Some of these laboratories contain prototype quantum computers, and 
therefore the computational problem of simulating the dynamics of these 
laboratories (and their many-qubit successors) must be BQP-hard. 
Moreover, one might make the argument that since nonrelativistic
quantum mechanics is a limiting case of quantum field theory and,
in principle, the laws of quantum mechanics permit the efficient
solution of the problems in BQP, it must be true that the problem
of simulating dynamics of quantum field theories is BQP-hard. 
What, then, can be learned from a derivation of BQP-hardness?

First, within a given quantum field theory, BQP-hardness depends on
the computational problem, which in physical terms corresponds to the 
set of observables and phenomena implementing the computation.
Furthermore,
a bigger goal is to study the whole space of quantum field theories in
terms of their computational power. 
In other words, it is interesting to investigate which quantum field 
theories (in arbitrary spacetime dimensions) give rise to classically 
tractable problems and which ones give rise to intractable problems. 
In particular, we aim to discover what features of a field theory
determine this division. For example, we wish to know if this property
is affected by integrability or quantum chaos. 
This paper takes a first step towards addressing some of these issues. 
In particular, we find that, with a sufficiently complex variation in 
the external fields, it is already hard to simulate a weakly-coupled 
quantum field theory that is in only one spatial dimension and is purely 
bosonic.


One of the central goals of computer science is to understand the
ultimate capabilities and limitations of computers. Since the seminal
works on quantum computation by Feynman and Deutsch in the 1980s, we
have known that this understanding cannot be achieved in isolation
from physics. The question thus becomes: What is the class of
computational problems solvable using polynomial resources within the
laws of physics that govern our universe? In essence, this work,
together with \cite{JLP14}, places matching upper and lower bounds on
the computational power of a universe described by $\phi^4$
theory. 
This represents a step
in a larger program of characterizing the computational power of the
Standard Model, which is the quantum field theory describing all
known physical phenomena other than gravity. Characterizing the
computational power of quantum gravity is a more distant goal.

\subsection{Proof sketch}

The essence of our BQP-hardness argument is to propose an architecture 
for a quantum computer in a $(1+1)$-dimensional universe governed by 
massive $\phi^4$ theory and then to show that it is capable of scalable, 
universal quantum computation. This is in some ways easier and in other
ways harder than designing a quantum computer architecture for
the real world. On the one hand, practical experimental limitations 
are not a concern; operations of arbitrary precision are allowed 
as long as the precision scales only polynomially with the problem size. 
On the other hand, the set of particles and interactions from which to 
construct the qubits and gates is much more limited.

In our BQP-hardness construction we choose our external field
$J_2(t,x)$ so that the nonrelativistic limit is a collection of
double-well potentials. A qubit is encoded by a double well containing
one particle. 
The logical-zero and logical-one states of the qubit can then be
represented by two states of the double well. For example, one
can choose the ground and first excited states 
of the double well as logical zero and one, respectively.
Another possible choice is particle occupation (in the ground state) 
of the left well for logical zero and the right well for logical one. 
We show that, in the nonrelativistic limit (with $J_1=0$), 
the effective $n$-particle Schr\"{o}dinger equation has the Hamiltonian
\begin{eqnarray}
H(t) &=& \sum_i \left( \frac{p_i^2}{2m} + \frac{J_2(t,x_i)}{m} \right) \nonumber \\
& + & \sum_{i<j} \left( \frac{\lambda}{4m^2}\big(1+\frac{\lambda}{4\pi m^2}\big)
 \delta(x_i-x_j) - \frac{\lambda^2}{32\pi m^3}
 \int_0^1 dy \frac{e^{-m|x_i-x_j|/\sqrt{y(1-y)}}}{\sqrt{y(1-y)}}
\right) \nonumber \\ & \qquad + \ldots \,.
\end{eqnarray}
By varying the source term $J_2(t,x)$ as a function of time, one can
move the potential wells. 
By moving the left and right wells of a single qubit closer together, one can
implement single-qubit gates through tunneling between the wells. By moving
the wells of neighboring qubits closer together, one can implement 
two-qubit gates through the inter-particle interactions. 
In this manner, we construct a universal set of quantum gates in 
Sec.~\ref{sec:gates}. 

An oscillatory $J_1(t,x)$ can create and destroy particles. 
This allows us to show that computing even the vacuum-to-vacuum 
transition probability is BQP-hard. To simulate a quantum
computation, we create a state in which 
each double well encoding a qubit is in its logical-zero state.
We show in Sec.~\ref{sec:prep} how we can prepare this state 
by starting with the vacuum state and then varying
the source term $J_1(t,x)$ sinusoidally in time. 
At the end, the time-reversed version of this process annihilates
the particles in the double wells. 
Thus, the $\ket{0\ldots0} \to \ket{0\ldots0}$ amplitude of a
quantum circuit corresponds to the vacuum-to-vacuum transition
amplitude, whereas other final states of the quantum circuit
contribute to non-vacuum final states of the quantum field theory's
dynamics.

The spatial volume used by this process is proportional to the number
of qubits, up to logarithmic factors, since the coupling between
wells decays exponentially with their spacing. 
The execution time of an individual quantum gate must scale as
$\widetilde{O}(\lambda^{-2})$, 
so that leakage errors out of the coding subspace are adiabatically
suppressed
(\sect{Subsec:Gate_Universality}). (The $\widetilde{O}$ suppresses
logarithmic factors, which come from adiabatic theorems with
exponential convergence.) The total duration of the
process is thus $\widetilde{O}(\lambda^{-2} D)$, where $D$ is the
depth of the original quantum circuit.

It is essential that the reduction can be carried out in polynomial 
time by classical computers. One potential concern is that
computing the function $J_2(t,x)$ corresponding to a specified quantum
circuit could be intractable.  We demonstrate that this is not the
case by giving examples of explicit constructions for $J_2(t,x)$ in
Appendices~\ref{xgate} and \ref{zgate}. For these constructions, we
assume that the coupling constant $\lambda$ is $O(1/G)$, where $G$ is
the number of gates in the simulated circuit, which allows us to use
low-order perturbation theory in $\lambda$ to analyze our two-qubit
gate to adequate accuracy. For such weak particle interactions, it
takes a time scaling like $G^2$ to execute a single two-qubit gate
adiabatically, and a time of order $G^2D$ for the simulation of a circuit 
with depth $D$. 

Let $T$ and $V$ be the duration and spatial volume on which $J_1$ and
$J_2$ have support. Then, by the Nyquist-Shannon theorem, it suffices
to use $O(T\omega V/\xi)$ real numbers to describe $J_1$ and $J_2$, where
$\omega$ and $\xi$ are the maximum frequency and minimum wavelength 
of $J_1$ and $J_2$.  Recall that 
$T = \widetilde{O}(\lambda^{-2} D) = \widetilde{O}(G^2 D)$. Since
the interaction between particles in separate wells falls off 
exponentially with separation, $V = \widetilde{O}(n)$, where $n$ is
the number of qubits. The wavelength $\xi$ is of order $1/m$, because the 
spacing and widths of wells that suppress unwanted tunneling are of order
$1/m$. The maximum oscillation frequency $\omega$ is of order $m$,
which occurs when an oscillatory $J_1$ term is used to excite
particles from the vacuum. The mass $m$ is taken to be a constant, not
varying asymptotically with problem size. Thus, the total number of
bits needed to specify $J_1$ and $J_2$ to adequate precision is 
$\widetilde{O}(nG^2D)$. 
This is important, because to show BQP-hardness one needs the reduction to
be computable classically in polynomial time and it must induce at
most a polynomial increase in the number of bits needed to describe
the problem instance.

Before we give the details of the proof starting with state preparation in the next section, we summarize the conditions for successful simulation and then discuss the level of rigor in our results.

\subsection{Conditions for successful simulation}
In this subsection, we collect together the various conditions need to design a $(1+1)$ dimensional $\phi^4$ theory to simulate an arbitrary quantum circuit.
\begin{enumerate}
\item Since we are interested in starting from vacuum in our simulation, the first condition ensures that a single particle populates the well with sufficient accuracy. Our state preparation method is to use adiabatic rapid passage where the time-dependent source term is modulated with the following envelope function.
\begin{equation}
f(t) = \left\{ \begin{array}{cl}
g \cos(\omega_0 t + B t^2/T) \,, & -T/2 \leq t \leq T/2 \,, \\
0 \,, & \textrm{otherwise} \,.
\end{array} 
\right.
\end{equation}
For successful creation of particles in wells, we need the parameters $g$, the  $B$ and $T$ to scale with $n$ as follows.
\begin{align}
g\sim 1/n^5, \quad \lambda \sim B \sim 1/n^4, \quad 
T \sim n^8.
\end{align}
This is derived in Section \ref{sec:prep} (Eq. \ref{eq:state-prep-scaling}).

\item To justify the perturbative treatment of the effective potential, and the non-relativistic approximation, we need $\lambda\sim J_2$. This is discussed in Section \ref{sec:gates}.

\item In our construction, we do not need to assume that the Hamiltonian is bounded. The adiabatic approximation is applied to the \emph{effective} Hamiltonian, which is bounded. In order to justify this approximation for the effective Hamiltonian, we need $\tau\sim \text{poly}(\log n)$, where $\tau$ is the time scale of adibatic evolution. This is described in detail in Section \ref{Subsec:Gate_Universality}.

\item When implementing gates, the wells could end up with two particles if the interaction time is too long. To prevent this double occupancy of potential wells, the gate run time $\sim \lambda^{-2}$. This is described in Section \ref{Subsec:Gate_Universality}.

\item In section \ref{sec:measurement}, we discuss the BQP completeness of the simulation of the vacuum-to-vacuum amplitude approximation problem. Although our BQP hardness result does not need the theory to be on a lattice, for BQP completeness, we use the algorithms of \cite{JLP12, JLP14} to prove that the decision version of the problem lies in BQP. For this result, we need a small enough lattice spacing to make lattice errors small as described in \cite{JLP12, JLP14}. In addition, we also need a sufficient number of runs that is polynomial in $n$, the input size, to estimate the vacuum-to-vacuum probability with small statistical errors.

\end{enumerate}

\subsection{Validity of perturbation theory}

For $\phi^4$ theory in two spacetime dimensions without sources, there exist proofs \cite{Eckmann:1976xa, Osterwalder:1975zn,Dimock:1976ys} that perturbation theory is asymptotic. However, this does not imply that including higher order terms reduces the error obtained from considering only a finite number of terms. This is because perturbation theory is not convergent and beyond a certain number of terms, the errors in approximation increase. Therefore, one only considers the first few terms in the perturbation expansion.

Unfortunately, there are no analogous results on the asymptotic nature of perturbation theory in the presence of sources. Due to this, the mathematical rigor of our simulation technique could be called into question when we use perturbation theory in the presence of time-varying source terms. We would like to state here explicitly that we assume perturbative approximations are reliable even in the presence of time-dependent sources. The other parts of our construction have explicit proofs either in the cited references or the appendices.

\section{State Preparation}
\label{sec:prep}

For the decision problem defined in the previous section,
our starting point is the vacuum state of the weakly interacting
$\phi^4$ theory with $J_2(x) = J_1(x) = 0$. First, we adiabatically turn on 
the static double-well potential (by turning on $J_2(x)$) to prepare the 
corresponding vacuum state. Next, we turn on an oscillatory $J_1(t,x)$ to
create a particle in the logical-zero state of each double well. 
The ground and first excited states of the double-well potential,
which are symmetric and antisymmetric superpositions over the two
wells, can serve as these logical states. 
It is also possible to choose a localized particle in the left or
right well, although these states are not eigenstates,
as long as the two wells are sufficiently separated; the energy
splitting between the two states is then exponentially suppressed
and, apart from accumulating a global phase, these states evolve
only exponentially slowly.

From the adiabatic theorem given in \cite{EH12}, it follows that one can
prepare the vacuum state of the static potential in a time of
$\widetilde{O}(1/(m-{\cal B})^2)$, where $m$ is the physical mass of the 
particles in the interacting theory and ${\cal B}$ is the binding energy 
of the well. Note that we cannot choose the binding energy of the well to 
be larger than the particle mass, because in this case 
the vacuum becomes unstable: it becomes energetically favorable to
create a particle occupying the well.

After creating the vacuum state for the system with nonzero $J_2$, we
wish to create exactly one particle in each double well. We do this by
applying an oscillating source term in the full interacting
theory. The idea is to ensure that the creation of one particle is on 
resonance, while the creation of more than one particle is
off resonance. Perhaps the simplest version of this procedure is to
use Rabi oscillation, in which we drive a transition to the
single-particle state using a $J_1(t,x)$ that oscillates on resonance
with the energy difference between this state and the vacuum. 

Because of the interaction term $\lambda \phi^4$ in the
Lagrangian, the system is anharmonic. While the energy to create
one particle in the ground state of a well is $m-{\cal B}$, the energy
to create two particles in the ground state is $2(m-{\cal B})-\delta$,
where $\delta$ is a binding energy arising from the inter-particle 
potential. 
Suppose we choose $J_1(t,x)$ to have the form
\begin{equation}
\label{straightsine}
J_1(t,x) = g \cos(\omega t) h(x) \,,
\end{equation}
where $h(x)$ is a function localized in the well (for example, a Gaussian),
and $g$ is a constant quantifying the strength of the source. If we
choose $\omega = m-{\cal B}$, then the source is on resonance for the 
transition from vacuum to the single-particle bound state, but off resonance 
for the transition from the single-particle bound state to the
two-particle bound state. 

Standard analysis of Rabi oscillation shows
that, for sufficiently weak $g$, the source $J_1(t,x)$ described in
\eq{straightsine} will drive oscillations between the vacuum and the
single-particle ground state $\ket{\psi_0}$ of the well with frequency
\begin{equation}
\label{rabi_freq}
\Omega = g \bra{\psi_0} \int dx h(x) \phi(x) \ket{\mathrm{vac}} \,.
\end{equation}
Thus, by applying the oscillating source for a time 
$\tau = \pi/2 \Omega$, 
one can drive a near-perfect transition to
the state $\ket{\psi_0}$. In principle, errors (that is,
excitations to higher-energy states) can be arbitrarily suppressed by making
$g$ smaller and $\tau$ correspondingly larger. Because we assume no
intermediate measurements in our scattering process, we cannot invoke
fault-tolerance constructions. Thus, each of the $n$ qubits must be
prepared with infidelity of $O(1/n)$.

The disadvantage of this construction is that, to prepare the 
state $\ket{\psi_0}$ with arbitrarily high fidelity, one 
needs arbitrarily precise knowledge of both the resonance frequency
$\omega = m-{\cal B}$ and the Rabi oscillation frequency $\Omega$ 
determined by the matrix element (\ref{rabi_freq}). Thus, we instead
apply a related scheme called adiabatic passage, which requires
only approximate knowledge of these quantities. 

In Sec.~\ref{sec:ARP}, we provide an overview of adiabatic passage.
Specifically, we present the theoretical description in the case of a 
two-level system. 
In the following subsections, we analyze the effect of source terms and
the application of adiabatic passage to our problem. 
In the familiar case of the free ($\lambda=0$) scalar theory without any
sources, one can simply go to Fourier space, express the Hamiltonian in terms 
of creation and annihilation operators for the Fourier modes, and thus obtain 
its spectrum.
The addition of a linearly coupled classical source of finite duration 
($J_1(x) \neq 0$ for a finite time) can also be treated straightforwardly.
In Sec.~\ref{sec:free+sources}, we present the analogous equations
when there is a quadratically coupled source ($J_2(x) \neq 0$). The
expressions make clear that $J_2(x)$ acts like a potential, with the
spectrum now having a discrete part as well as a continuum.
One can furthermore see that the choice of $J_1(x)$ determines the 
probabilities of various particle types being created.
Next, in Sec.~\ref{sec:interact}, we examine the interacting $\phi^4$
theory with both sources. In particular, we consider $J_1(x)$ with 
a time dependence that implements adiabatic passage.
Calculation of the Fourier spectrum of such a function of time reveals that
one can suppress transitions to states that are not in resonance with
the desired transition and whose transition frequency is not a multiple of the 
desired transition frequency.
Hence, one can suppress the production of multiple particles in the 
discrete part of the spectrum, since it is anharmonic. 
We show that the production of unwanted unbound particles can also be 
suppressed. Thus, we obtain a set of necessary conditions for successful 
state preparation. The parameter scalings that satisfy these conditions
determine the time required.

\subsection{Adiabatic Passage}
\label{sec:ARP}

Adiabatic passage is an experimental technique for driving transitions 
between eigenstates. 
Instead of applying a sinusoidal driving term tuned precisely to the 
desired transition frequency, as in Rabi oscillation, one applies a
driving term whose frequency sweeps across this resonance. For our
purposes, the advantage of this technique is that excited-state
preparation of  arbitrarily high fidelity is achievable with only
limited knowledge of the relevant transition frequency and
matrix element. 
The theoretical description of such a coupled system is summarized below. 

\subsubsection{Effective Hamiltonian in the Rotating Frame}
Consider a two-level system with energy splitting $\omega_0>0$, where the 
transition between the ground and excited states is being driven by a source 
with frequency $\omega = \omega_0 +\Delta$; we say that the driving field is 
``detuned'' from resonance by $\Delta$. This system satisfies the 
Schr\"odinger equation
\begin{equation}
\frac{d}{dt} |\psi\rangle = -iH(t)|\psi\rangle,
\end{equation}
with the time-dependent Hamiltonian 
\begin{equation}\label{eq:time-dependent-Ham}
H(t) =\left[
\begin{array}{cc}
  0 &  e^{i\omega t}\Omega/2  \\  e^{-i\omega t}\Omega /2& \omega_0      
\end{array}
\right]\,.
\end{equation}
When expressed in terms of a ``rotating frame'', 
Eq.~(\ref{eq:time-dependent-Ham}) becomes
\begin{equation}
\frac{d}{dt} |\varphi\rangle = -iH_{\rm eff}|\varphi\rangle \,.
\end{equation}
Here,
\begin{equation}
|\psi \rangle = \left[
\begin{array}{cc}
 1&0   \\ 0&e^{-i\omega t}      
\end{array}
\right]
|\varphi\rangle \,,
\end{equation}
and the effective Hamiltonian is 
\begin{equation}
\label{eq:Hamiltonian-theta}
\begin{array}{rcl}
H_{\rm eff}  & =  &
\left[\begin{array}{cc}
 0 &  \Omega/2  \\  \Omega/2 & -\Delta      
\end{array} \right] \vspace{5pt} \\
& = &
\left[\begin{array}{cc}
  -\Delta/2 &  0 \\  0 & -\Delta/2   
\end{array}\right] \vspace{5pt}
 + \frac{1}{2} \sqrt{\Omega^2+\Delta^2}\left[
\begin{array}{cc}
 - \cos 2\theta &  \sin 2\theta  \\  \sin 2\theta & \cos 2\theta    
\end{array}
\right] \,,
\end{array}
\end{equation}
where $\tan 2\theta = -\Omega/\Delta$ and $0\le \theta \le \pi/2$. 
The eigenstates of this effective Hamiltonian are
\begin{eqnarray}
\ket{+} & = & \sin\theta |g\rangle + \cos\theta |e\rangle\,, \label{eq:ham-eigenstates} \\
\ket{-} & = & \cos\theta |g\rangle -\sin\theta |e\rangle\,,  \label{eq:estate}
\end{eqnarray}
with eigenvalues $-\frac{1}{2}\Delta \pm \frac{1}{2}\sqrt{\Omega^2+\Delta^2}$, 
respectively. Here, $|g\rangle$ and $|e\rangle$ are the ground and excited 
states of the undriven Hamiltonian with $\Omega = 0$ and $\omega =0$.

\begin{figure}[htb]
\begin{center}
\includegraphics[width=0.65\textwidth]{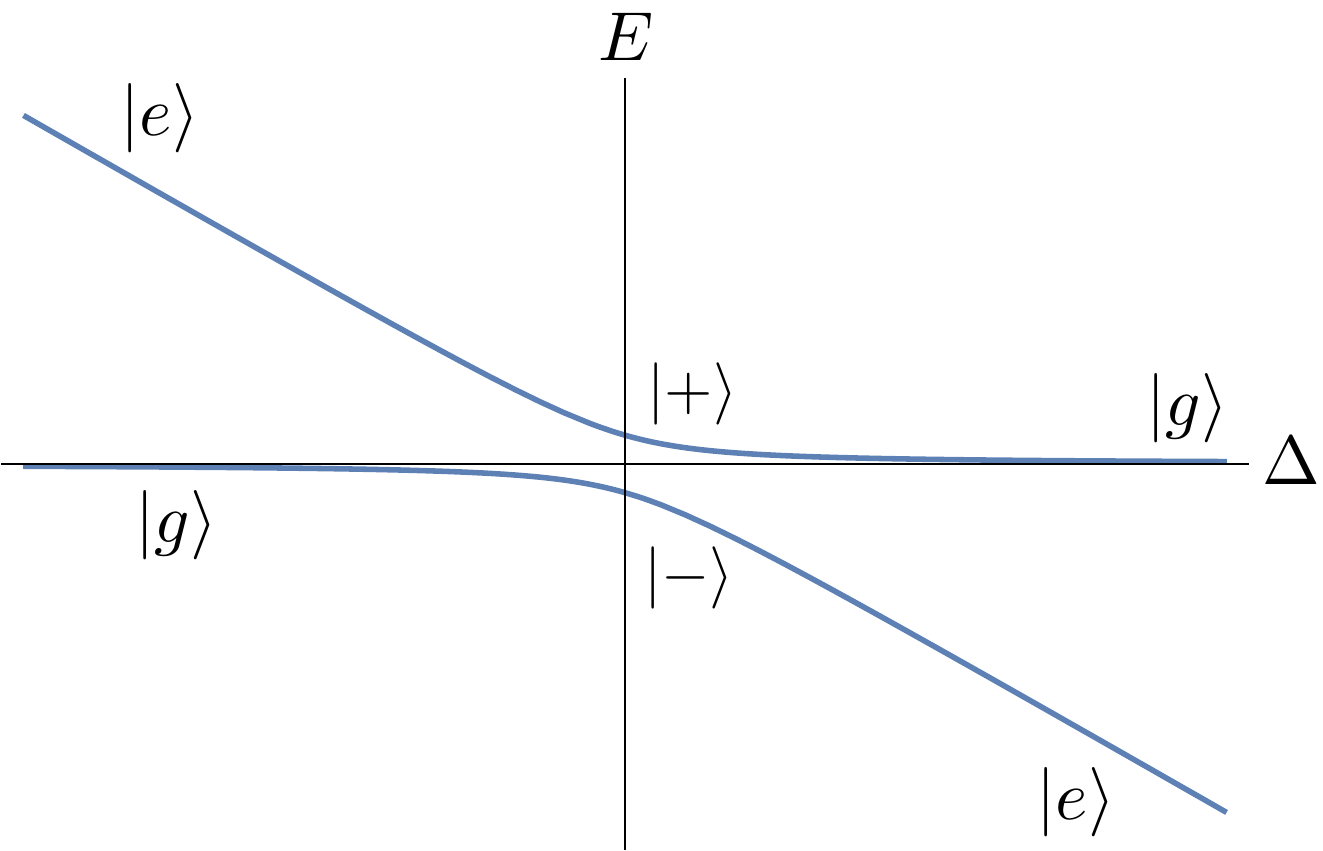}
\caption{Avoided crossing of energy levels. As the detuning $\Delta$
sweeps adiabatically through zero, the eigenstate $\ket{-}$ changes
from the uncoupled ground state $\ket{g}$ to the uncoupled excited state
$\ket{e}$.}
\label{fig:arp}
\end{center}
\end{figure}

In adiabatic passage, the detuning $\Delta$ sweeps through zero. 
Well below resonance 
(large negative detuning $\Delta$, $\theta\approx 0$), the lower-energy 
eigenstate of $H_{\rm eff}$ is $|-\rangle \approx |g\rangle$, whereas 
well above resonance (large positive $\Delta$, $\theta \approx \pi/2$) 
we have $|-\rangle \approx -|e\rangle$. If the sweep is slow enough, the 
system is unlikely to be excited across the minimal energy gap $\Omega$ 
of the effective Hamiltonian, and it evolves adiabatically from the ground 
state $|g\rangle$ to the excited state $|e\rangle$. 
%
%
(See Fig.~\ref{fig:arp}.)

We emphasize that, even if the off-diagonal term in $H(t)$  is small, we 
cannot analyze adiabatic passage by treating this term in time-dependent 
perturbation theory; for adiabatic passage to succeed, the off-diagonal 
driving term must be turned on for long enough that its effects 
are not perturbatively small. Correspondingly, in a field-theory setting, the 
probability of successful particle creation by adiabatic passage cannot be 
computed by summing a finite number of Feynman diagrams: instead, 
resummation of an infinite class of diagrams is required. Our strategy will be 
to justify approximating the field theory problem by the two-level system just 
described, and then to compute the success probability within that two-level 
approximation. What must be shown is that terms in the Hamiltonian coupling 
these two energy levels to other energy levels can be safely neglected. This 
issue arises in any treatment of adiabatic passage between two energy levels 
of a multilevel system. 

\subsubsection{Rotating-Wave Approximation}

In Eq.~(\ref{straightsine}), the source term $J_1(t,x)$ is proportional to 
$\cos(\omega t)$, whereas in Eq.~(\ref{eq:time-dependent-Ham}) we included only 
off-diagonal terms with small detuning, neglecting counter-rotating terms, which
are far from resonance. Intuitively, these terms, which oscillate rapidly in 
the rotating frame, have effects that nearly average away. Ignoring the 
counter-rotating terms in the Hamiltonian is called the {\em rotating-wave 
approximation}.

In the rotating frame, the counter-rotating part of the Hamiltonian is
\begin{equation}
H_{\rm eff}'(t) = \left[
\begin{array}{cc}
  0 &  e^{-2i\omega t}\Omega/2  \\  e^{2i\omega t}\Omega /2&  0
\end{array}
\right]\,.
\end{equation}
To justify the rotating wave approximation, we first express the Schr\"odinger 
equation as an integral equation, namely,
\begin{equation}
|\varphi(T)\rangle = |\varphi(0)\rangle - i \int_0^T dt H_{\rm eff}(t) |\varphi(t)\rangle,
\end{equation}
and note that the error ignoring the counter-rotating term introduces is 
\begin{equation}
|\epsilon\rangle=   i \int_0^T dt H_{\rm eff}'(t) |\varphi(t)\rangle,
\end{equation}
which after an integration by parts becomes
\begin{equation}
|\epsilon\rangle = \frac{\Omega}{4\omega}\left( \left[\begin{array}{cc}
  0 & - e^{-2i\omega t}  \\  e^{2i\omega t}&  0
\end{array}\right]|\varphi(t)\rangle \right)_0^T
 + i\frac{\Omega}{4\omega} \int_0^T dt \left[\begin{array}{cc}
  0 & - e^{-2i\omega t}  \\  e^{2i\omega t}&  0
\end{array}\right]
H_{\rm eff}(t) |\varphi(t)\rangle.
\end{equation}
We can therefore bound the error using
\begin{eqnarray}
\epsilon := \||\epsilon\rangle\| & \le & \frac{\Omega}{2\omega} + \frac{\Omega T}{4\omega} \max_{t\in[0,t]} \|H_{\rm eff}(t) \| \nonumber \\
& \le & \frac{\Omega}{2\omega} + \frac{\Omega T}{4\omega}\left(\Delta + \Omega\right). \label{eq:rotating-wave-error}
\end{eqnarray}
We can use a similar argument to bound the contribution from rapidly oscillating on-diagonal terms in the Hamiltonian (which are also rapidly oscillating in the rotating frame).

\subsubsection{Conditions for Successful Adiabatic Passage}

To be concrete, consider the two-level Hamiltonian
\begin{equation}\label{eq:ham-h(t)}
H(t) =\left[
\begin{array}{cc}
  0 &  h(t) \\ h(t)& \omega      
\end{array}
\right]\,,
\end{equation}
where the off-diagonal driving term has the time dependence
\begin{equation}
h(t) = \begin{cases}
\Omega \cos (\tilde \omega(t) t), & -T/2 \le t \le T/2,\\
0, & {\rm otherwise},
\end{cases}
\end{equation}
and
\begin{equation}\label{eq:delta-sweep}
\tilde \omega(t) = \omega_0 + \Delta(t), \quad \Delta(t) = Bt/T.
\end{equation}
In other words, this source term is turned on for a total time $T$, during 
which the detuning ramps linearly in time from $-B/2$ to $B/2$. We refer to 
$B$ as the (circular) frequency bandwidth of the time-dependent source. 
In the rotating-wave approximation, the effective Hamiltonian has the form 
of Eq.~(\ref{eq:Hamiltonian-theta}), with $\Delta$ given by 
Eq.~(\ref{eq:delta-sweep}). We wish to find sufficient conditions for passage 
from the initial state $|g\rangle$ to the final state $|e\rangle$ to occur 
with a small error $\epsilon$. 

First, when the source turns on suddenly at $t= -T/2$, we want the initial 
state $|g\rangle$ to be close to the eigenstate $|-\rangle$ in 
Eq.~(\ref{eq:ham-eigenstates}), and when the source turns off at $t=T/2$ we 
want $|-\rangle$ to be close to $|e\rangle$. To ensure that the error due to 
misalignment of $|-\rangle$ with the initial and target states is sufficiently 
small, we impose the condition
\begin{equation}\label{eq:condition-omega-B}
\Omega/B = O(\epsilon).
\end{equation}

Second, we need the sweep of the detuning to be slow enough for the evolution 
to be adiabatic. The effective Hamiltonian (in the rotating-wave approximation)
obeys
\begin{equation}
\left\| \frac{d}{ds} H_{\rm eff}(s)\right\|  = \left|\frac{d}{ds} \Delta(s) \right|= B,
\end{equation}
where $s = t/T$,
and its minimum gap is $\gamma=\Omega$; therefore, by the adiabatic 
theorem~\cite{Ruskai}, the  error due to a diabatic transition will be 
$O(\epsilon)$ provided
\begin{equation}\label{eq:condition-T-long}
 \frac{1}{T\gamma^3}\left\| \frac{d}{ds} H_{\rm eff}(s)\right\|^2= \frac{B^2}{T\Omega^3}=O(\epsilon).
\end{equation} 

Third, for the corrections to the rotating-wave approximation bounded in 
Eq.~(\ref{eq:rotating-wave-error}) to be small, we impose the conditions
\begin{equation}\label{eq:condition-T-short}
\Omega / \omega_0 = O(\epsilon), \quad \Omega BT/\omega_0 = O(\epsilon).
\end{equation}
Note that, while we require $T$ to be small enough to justify the rotating-wave
approximation, it must also be large enough to ensure adiabaticity during the 
sweep.

The conditions listed so far already arise in the analysis of the two-level 
system, Eq.~(\ref{eq:ham-h(t)}). We need to impose further conditions to 
ensure that the amplitude is small for transitions from these two levels to 
other states and thereby justify the two-level approximation.
With that purpose in mind, we now discuss particle creation by a 
time-dependent source in a field-theory context, first in a free theory and 
then in an interacting theory. 


\subsection{Free Theory with Sources}
\label{sec:free+sources}

Consider first the free theory with a static quadratically coupled
source.
One can analyze the effect of the source through a straightforward
generalization of standard sourceless free-theory calculations.
The Hamiltonian can be expressed in terms of creation and annihilation
operators $a_l^\dagger$ and $a_l$
as
\begin{equation}
	H = \sumint{l} \omega_l \big( a_l^\dagger a_l 
		+ \frac{1}{2}[a_l,a_l^\dagger]\big) \,,
\end{equation}
where
%
%
\begin{equation}
	(-\partial_x^2 + m^2 + 2J_2(x))\psi_l(x) = \omega_l^2 \psi_l(x) 
 \label{eq:SchrEq}
\end{equation}
and
\begin{equation}
        \phi(t,x) =\sumint{l} \frac{1}{\sqrt{2\omega_l}} 
                \big( a_l \psi_l^*(x)e^{-i\omega_lt} 
                + a_l^\dagger(x) \psi_l(x)e^{i\omega_lt} \big)
        \,.
\end{equation}
In other words, $\psi_l(x)$ and $\omega_l^2$ are the energy 
eigenfunctions and eigenvalues of a Schr\"{o}dinger equation with potential 
$m^2(x) = m^2 + 2J_2(x)$. 
Here, $\sumint{}$ indicates a sum over the discrete part plus an
integral over the continuous part of the spectrum. 

Thus, the spectrum consists of particles associated with the
solution of the Schr\"{o}dinger equation with a potential determined by 
the source term $J_2(x)\phi^2$. 

\medskip

Now consider turning on a linearly coupled source $J_1(t,x)\phi(t,x)$ for a
finite time.
Let
\begin{equation}
\tilde{J}_1(\omega_l,l) = \int d^2y \psi_l(y^1) e^{i \omega_l y^0} J(y^0,y^1) 
\,.
\end{equation}
(In the special case $J_2=0$, the $\psi_l$ are simply plane-wave solutions
with  $\omega_{p}^2 = {p}^2 + m^2$, 
and $\tilde{J}_1$ is then the Fourier transform of $J_1(x)$.)
Using the equation of motion and the retarded Green's function, one finds
that
\begin{equation}
        H = \sumint{l} \omega_l \big( b_l^\dagger b_l 
                + \frac{1}{2}[b_l,b_l^\dagger]\big) \,,
\end{equation}
where
\begin{equation}
b_l = a_l + \frac{i}{\sqrt{2\omega_l}}\tilde{J}_1(\omega_l,l)
\,.
\end{equation}
%
The probabilities of no particles being created, $P(0)$, and
a single $k$-type particle being produced, $P(k)$, are
\begin{eqnarray}\label{eq:P}
P(0) & = & |A(0)|^2 = \exp \left[- \sumint{l} \frac{1}{2\omega_l}
        |\tilde{J}_1(\omega_l,l)|^2 \right] \\
P(k) & = & |A(k)|^2
       = |\tilde{J}_1(\omega_k,k)|^2 
		    \exp \left[- \sumint{l} \frac{1}{2\omega_l}
        |\tilde{J}_1(\omega_l,l)|^2 \right] \, . \label{eq:P2}
\end{eqnarray}
Production of $n$ particles can also occur, with the probability given 
by the Poisson distribution. 
Thus, the non-interacting theory does not allow one adequately to suppress
creation of more than one particle from the vacuum state. 


\subsection{Interacting Theory with Sources}
\label{sec:interact}
Suppose that the time-dependent source term in the Hamiltonian density is 
$J_1(t,x)\phi(t,x)$, where
\begin{equation}
J_1(t,x) = f(t)h(x) 
\end{equation}
and
\begin{equation}
f(t) = \left\{ \begin{array}{cl}
g \cos(\omega_0 t + B t^2/T) \,, & -T/2 \leq t \leq T/2 \,, \\
0 \,, & \textrm{otherwise} \,.
\end{array} 
\right.
\end{equation}
If the two-level approximation is justified, then the analysis of adiabatic 
passage in Sec.~\ref{sec:ARP} applies, with the Rabi frequency $\Omega$ 
given by Eq.~(\ref{rabi_freq}). 

In Appendix \ref{app:fourier}, we analyze the properties of ${\cal F}(\omega)$,
the Fourier transform of $f(t)$. There, it is shown that $\cal F(\omega)$ is 
approximately constant for frequencies in a band of width $B$ centered at 
$\omega_0$: 
\begin{equation}
{\cal F}(\omega) \approx g\sqrt{T/B}, \quad 
\omega_0- B/2 < \omega < \omega_0 + B/2.
\end{equation}
Outside this band, ${\cal F}(\omega)$ is much smaller: for $\delta$ scaling 
like $\sqrt{B/T}$, we have
\begin{equation}
{\cal F}(\omega) = O( g/B), \quad | \omega - \omega_0| \ge B/2 + \delta.
\end{equation}
Thus, for $BT \gg 1$, ${\cal F}$ is well approximated by a rectangular function in frequency space, supported on the interval $|\omega - \omega_0| \le  B/2$.

Now consider the effect of the $\lambda \phi^4$ interaction. Recall 
that we cannot use perturbation theory to analyze the production 
of a single bound particle by adiabatic passage, instead needing the
nonperturbative analysis of the two-level effective Hamiltonian described in 
Sec.~\ref{sec:ARP}. Nevertheless, we can use perturbation theory to bound 
the error arising from unwanted transitions. 

One process arising in the free theory that we need to consider is the 
production of more than one particle in the potential well. In the interacting 
theory, the energy of a two-particle state is shifted from twice the energy of 
a one-particle state by $O(\lambda)$, because of the interaction between 
particles. The amplitude for a transition that changes the energy by 
$\omega$ scales like ${\cal F}(\omega)$. To suppress production of multiple 
particles, we need the coupling $\lambda$ to be large enough to shift 
the energy of the transition from one particle to two particles outside the 
band where ${\cal F}(\omega)$ is large. We therefore require
\begin{equation}\label{eq:condition-B-lambda}
B = O(\lambda).
\end{equation}
There is still a contribution to the error from the amplitude for the 
$1\to 2$ transition driven by the source outside that frequency window; 
we therefore impose
\begin{equation}\label{eq:condition-g-B}
g/B = O(\epsilon),
\end{equation}
where $\epsilon$ is the error. 

In contrast with the two-particle bound state (whose energy is shifted
away from $2 \omega$), there is a single-particle unbound state with
momentum $p_{2 \omega}$ such that the total energy is exactly $2 \omega$.
The transition to this state from the single-particle state of energy
$\omega$ is on resonance, that is, it is not suppressed by the decaying
tail of $\mathcal{F}(\omega)$.
However, one can suppress this unwanted excitation by judiciously choosing
the spatial profile $h(x)$ so that the matrix element of the term
$\int dx h(x) \phi(x)$ coupling the single-particle bound state to
the single-particle momentum-$p_{2 \omega}$ unbound state is small. By
computing the spatial profile $\psi_0(x)$ of the mode corresponding to the
single-particle bound state, one can in fact ensure that this matrix element
is precisely zero. For $\lambda = 0$, one can solve for $\psi_0(x)$ exactly
if the double wells defined by $J_2(x)$ are chosen from among the double-well
potentials with known exact solutions. Then, as $\lambda \to 0$ this
approximation becomes parametrically more precise.

\begin{figure}[htb]
\begin{center}
\includegraphics[width=0.2\textwidth]{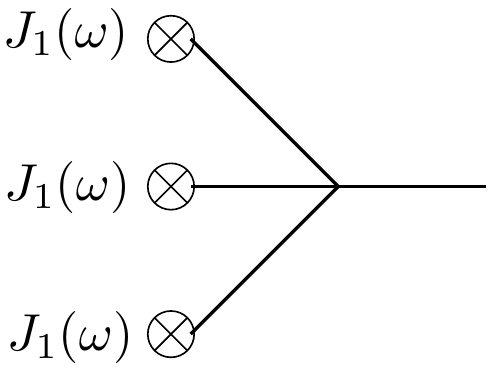}
\caption{Example of a connected diagram, suppressed by 
$(\lambda/m^2)(J_1/m^2)^3$, in which three insertions of
the source each with frequency $\omega$ produce a single particle 
with frequency $3\omega$. This diagram is the most dangerous because other diagrams are suppressed by additional powers of the small quantities, namely, $\lambda$ and $(J_1/m^2)$.}
\label{fig:feyndiag}
\end{center}
\end{figure}

We also need to take into account the production of unwanted states that are on 
resonance with multiples of the source frequency, including the production 
of unbound particles in the continuum. There are connected Feynman diagrams, 
higher order in $\lambda$, in which $k$ $J_1$ insertions, each with frequency 
in the band where ${\cal F}(\omega)$ is large, combine to produce one or more 
particles with total energy of approximately $k \omega_0$. 
An example is shown in Fig.~\ref{fig:feyndiag}. There are also additional 
tree-level diagrams suppressed by more powers of $(\lambda/m^2)( J_1/m^2)^2$, 
as well as $O(\lambda)$ loop corrections. The most dangerous diagram, the one 
shown in Fig.~\ref{fig:feyndiag}, scales like $\lambda{\cal F}(\omega)^3$; 
to ensure that it is adequately small, we impose
\begin{equation}\label{eq:condition-g-B-T}
\lambda \left(g \sqrt{T/B}\right)^3 
\sim \frac{\left(g \sqrt{T}\right)^3}{\sqrt{B}}= O(\epsilon),
\end{equation}
where we have used $\lambda \sim B$. Other diagrams are not problematic because they are suppressed by powers $\lambda$ and $(J_1/m^2)$, which are small quantities.

In Eqs.~(\ref{eq:condition-omega-B}), (\ref{eq:condition-T-long}), 
(\ref{eq:condition-T-short}), (\ref{eq:condition-B-lambda}), 
(\ref{eq:condition-g-B}), and (\ref{eq:condition-g-B-T}), we have now 
enumerated a set of sufficient conditions to ensure that our state preparation 
by adiabatic passage succeeds with a small error $\epsilon$ 
(where we recall that $\Omega \sim g$). We just need to check that all of 
these conditions can be satisfied simultaneously. One can verify that all 
conditions are satisfied by choosing parameters to scale with $\epsilon$ as 
follows:
\begin{equation}\label{eq:state-prep-scaling}
g\sim \epsilon^5, \quad \lambda \sim B \sim \epsilon^4, \quad 
T \sim 1/\epsilon^8.
\end{equation}
If we wish to simulate an $n$-qubit circuit accurately, the error $\epsilon$ 
in the preparation of each qubit should scale like $1/n$. We conclude that the 
state preparation can be achieved in a time of order $n^8$, 
if all the qubits are prepared in parallel. 

However, as we discuss in Sec.~\ref{sec:gates}, in the simulation of a circuit 
with $G$ gates, we choose $\lambda \sim 1/G$ to ensure that the action of our 
entangling two-qubit gate can be computed classically both accurately and 
efficiently. For a deep circuit, the requirement $\lambda \sim 1/G$ is more 
stringent than the condition $\lambda \sim 1/n^4$ implied by 
Eq.~(\ref{eq:state-prep-scaling}), and therefore $g$ must be correspondingly 
smaller as well, and $T$ correspondingly larger. We conclude that the state 
preparation can be achieved in the time
\begin{equation}
T \sim \max(n^8, G^{2})\,.
\end{equation}
If $G$ is much larger than $n^4$, the  corresponding state-preparation error 
of order $G^{-1/4}$ is actually much smaller than is needed for an accurate 
simulation. 


\section{Quantum Gates}
\label{sec:gates}

In this section, we describe how one can perform universal quantum 
computation. We first discuss how what is perhaps the most obvious attempt
does not work and then explain how one can overcome this difficulty. 
One might naively try to choose the encoding and then implement each 
gate from a particular universal gate set. This set must include a
two-qubit gate, which one would try to realize by bringing the qubit
double wells closer to each other.
For instance, to implement a controlled-phase gate with a dual-rail
encoding, one would decrease the separation of the logical-one wells, 
with the intention that the interactions between the particles would 
implement the operation.
The problem with this idea is that tunneling of particles between
double wells can occur. Tunneling leads to states not in the computational
subspace; in short, the qubit encoding is destroyed.
In Sec.~\ref{Subsec:Gate_Times}, we demonstrate that there is no regime
in which the particle interaction is parametrically larger than
the tunneling between double wells.

Instead, we can achieve universality 
by using unitary operations within the larger space of all six configurations
in which four wells are occupied by two particles without double
occupations. (Transitions to doubly occupied states are suppressed by
adiabaticity.) In Sec.~\ref{Subsec:Gate_Universality}, we describe how to 
realize unitary transformations within this larger space that closely 
approximate entangling two-qubit gates acting on the computational subspace.
Our analysis uses adiabatic theorems, which show how
slowly one must perform the operations in order to implement a gate
with a specified precision. 


Before we present the details, let us examine quantitatively
the effects of the static source and particle interactions.
Consider the Lagrangian 
after $J_1$ has been used for state preparation and turned off:
\begin{equation}
{\cal L} = \frac{1}{2}\partial_\mu\phi\partial^\mu\phi
           - \frac{1}{2} m^2 \phi^2 - \frac{1}{4 !}\lambda\phi^4
	   - J_2\phi^2 \,.
\end{equation}
One can obtain the tree-level (lowest-order in $\lambda$)  
nonrelativistic Lagrangian as follows.
Let
\begin{equation}
\phi \equiv \frac{1}{\sqrt{2m}}(e^{-imt}\psi + e^{imt}\psi^*) \,.
\end{equation}
Then,
\begin{equation}
{\cal L}_{NR} = i \dot{\psi}\psi^* + \frac{1}{2m} \psi^*\nabla^2\psi
	   - \frac{J_2}{m} \psi^*\psi 
	   - \frac{\lambda}{16m^2}(\psi^*\psi)^2\,.
\end{equation}
Thus,
$\pi = \partial{\cal L}_{NR}/\partial\dot{\psi} = i\psi^*$,
and the nonrelativistic Hamiltonian
${\cal H}_{NR} = \pi \dot{\psi} - {\cal L}_{NR}$
is
\begin{equation}
{\cal H}_{NR} = -\frac{1}{2m} \psi^* \nabla^2\psi 
	+ \frac{J_2}{m} \psi^*\psi 
	+ \frac{\lambda}{16m^2}(\psi^*\psi)^2\,.
\end{equation}
Since $p = -i \nabla$, the lowest-order Schr\"{o}dinger-picture 
Hamiltonian is
\begin{equation}
{\cal H}_{NR} = \sum_i \frac{p_i^2}{2m} + \sum_i \frac{J_2(x_i)}{m}
+ \frac{\lambda}{4m^2}\sum_{i<j}\delta(x_i-x_j) \, ,
\end{equation}
where $x_i$ denotes the position of particle $i$.
Including the ${\cal O}(\lambda^2)$ term (see \cite{JLP14}), 
we obtain
\begin{eqnarray}
{\cal H}_{NR} & = & \sum_i \frac{p_i^2}{2m} + \sum_i \frac{J_2(x_i)}{m} 
 + \frac{\lambda}{4m^2}\Big(1+\frac{\lambda}{4\pi m^2}\Big)
       \sum_{i<j}\delta(x_i-x_j) \nonumber \\
& - & \frac{\lambda^2}{32\pi m^3} \sum_{i<j}
 \int_0^1 dy \frac{e^{-mr_{ij}/\sqrt{y(1-y)}}}{\sqrt{y(1-y)}}
\label{eq:HNR}
\end{eqnarray}
where $r_{ij}$ is the distance between particles $i$ and $j$. We see that 
the static source $J_2$ induces a nonrelativistic effective potential 
$V(x) = J_2(x)/m$. Equation~(\ref{eq:HNR}) shows only the lowest-order terms 
in $p^2$ and $J_2$; higher-order terms can also be efficiently computed. 

We choose $J_2 \sim \lambda$, which ensures that both the binding energy 
and the kinetic  energy of a particle in a potential well scales like 
$\lambda$. As we shall explain in Sec.~\ref{Subsec:Gate_Universality}, it 
takes a time of order $\lambda^{-2}$ to execute our two-qubit entangling gate. 
Therefore, by including all terms up to order $\lambda^2$ in the effective 
Hamiltonian, we can compute the action of the gate up to an $O(\lambda)$ 
error. For this purpose, there are some contributions we need to 
include beyond what is shown in Eq.~(\ref{eq:HNR}). One is the first 
relativistic correction to the kinetic energy, namely $-\sum_i p_i^4/8m^3$. 
In addition, there are $O(\lambda J_2)$ terms arising from Feynman diagrams 
with a single point interaction and also a $J_2$ insertion on one of the four 
external legs. For a specific choice of $J_2$, these diagrams can be computed 
numerically. 

To justify using perturbation theory up to $O(\lambda^2)$ for the purpose 
of computing the action of the gate, we recall that perturbation theory is 
provably asymptotic in $\phi^4$ theory (without sources) in two spacetime 
dimensions (and, more generally, in two-dimensional theories in which the 
interaction is a polynomial in $\phi$)~\cite{Eckmann:1976xa, 
Osterwalder:1975zn,Dimock:1976ys}: 
when scattering matrix elements are computed in perturbation theory to 
$N\th$ order in $\lambda$, the error is $O(\lambda^{N+1})$ as 
$\lambda \to 0$.

\subsection{Gate Times}\label{Subsec:Gate_Times}

The inter-particle potential is created by the 
scattering of two particles. 
For $\phi^4$ theory, in which there is only one 
type of particle (a massive scalar), the potential has an $O(\lambda)$ 
repulsive contact term (that is, a term proportional to the delta function) 
and an $O(\lambda^2)$ exponentially decaying attractive term arising from 
the exchange of two massive particles. 

To analyze the entangling gate between a pair of dual-rail-encoded qubits, 
we consider the interaction between two particles, each confined to a 
potential well, where the wells are widely separated. The leading 
contribution to the phase shift comes from the contact interaction and can be 
efficiently computed. In a circuit with $G$ gates, to ensure a small error, 
we wish to specify the action of each gate to infidelity $O(1/G)$. For this 
reason, we choose $\lambda = O(1/G)$, so that corrections to the phase shift
that are higher order in $\lambda$ can be neglected. 

In addition to this phase shift, which occurs when both potential wells are 
occupied by particles, we need to consider the tunneling between wells that 
occurs when one of the two wells is unoccupied, which can also be efficiently 
computed. For a potential barrier with height $V$ and width $\ell$, the 
tunneling matrix element for a particle with energy $E$ scales like 
${\cal W} =\exp(-\ell\sqrt{2m(V-E)})$. The interaction energy of two particles 
of energy $E$ separated by the barrier, due to the overlap of their wave 
functions and the contact interaction, scales like 
$(\lambda/m^2)\times  {\cal W}^2$ and is therefore parametrically small 
compared with the tunneling matrix element when $\lambda$ is small. Thus, the 
time needed to generate a large phase shift is large compared with the 
tunneling time, and tunneling cannot be ignored during the execution of a 
two-qubit entangling gate. 

\begin{figure}[htb]
\begin{center}
\includegraphics[width=0.7\textwidth]{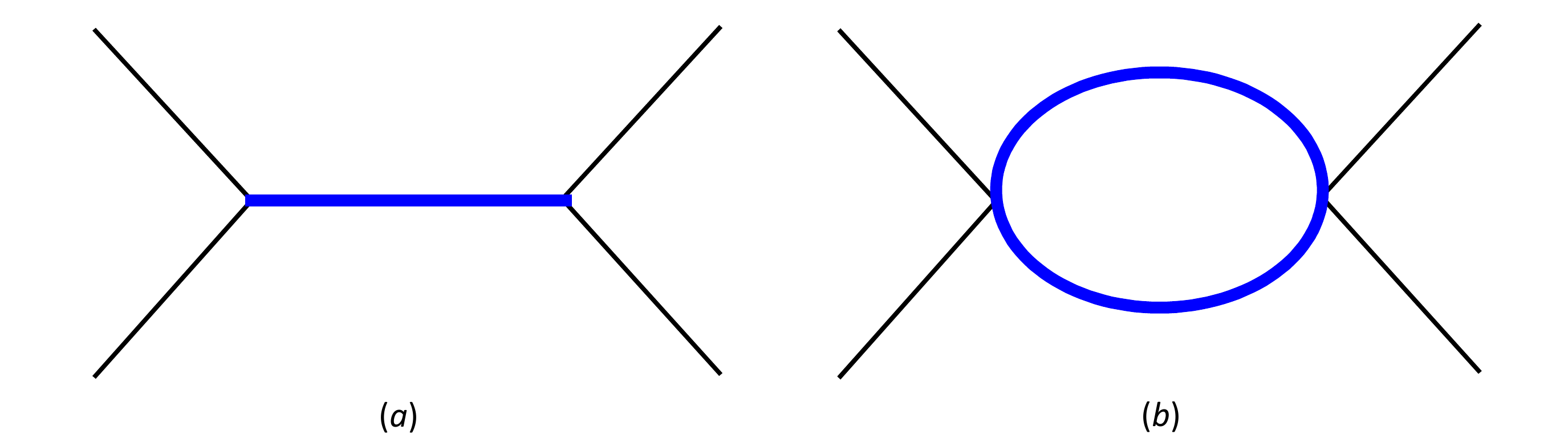}
\caption{Feynman diagrams contributing to two-particle scattering at order 
$\lambda^2$ for ($a$) a $\lambda\phi^3$ interaction and ($b$) a 
$\lambda \phi^4$ interaction. 
The heavy internal lines are dressed propagators. }
\label{fig:feynman-potential}
\end{center}
\end{figure}

Let us verify that the contributions to the interaction energy that are 
higher order in $\lambda$ can be safely neglected. Feynman diagrams 
corresponding to the leading contributions to $2\to 2$ scattering from 
particle exchange are shown in Fig.~\ref{fig:feynman-potential}, for both 
$\phi^3$ and $\phi^4$ interaction terms. In  
Fig.~\ref{fig:feynman-potential}$a$ 
a single particle is exchanged; the internal line in the diagram is the 
dressed propagator
\begin{equation}
\bar{\Delta}_{x_1,x_2}=-\left(\frac{1}{-\frac{d^2}{dx^2}+m^2+2J_2(x)}\right)_{x_1,x_2}\,,
\end{equation}
which includes the effects of the $J_2\phi^2$ source term.

To be concrete, consider the exactly solvable case 
in which $J_2$ is a static square barrier of
width $\ell$ and height $mV$. 
We wish to compute the Green's function
\begin{equation}
G(x_1,x_2;z)= \left(H - z \right)^{-1}_{x_1,x_2},
\end{equation}
where
\begin{equation}
H = -\frac{1}{2m}\frac{d^2}{dx^2} + V(x), 
\end{equation}
$J_2 = mV$, and $z= -m/2$. We can evaluate $G(x_1,x_2;z)$ in terms of the 
Wronskian (see \cite{Baltin}). 

Let $u_L(x;z)$ and $u_R(x;z)$ be solutions to the eigenvalue equation
\begin{equation}
Hu(x;z)=zu(x;z)\,,
\end{equation}
which approach zero as $x\to -\infty$ and $x\to +\infty$ respectively. 
The Wronskian is then
\begin{equation}
W(z)=u_L^\prime(x;z)u_R(x;z)-u_L(x;z)u_R^\prime(x;z)\,.
\end{equation}
It can be shown that $dW/dx = 0$, so the Wronskian is independent of $x$
and depends only on the eigenvalue $z$. 
The Green's function can be written in
terms of the Wronskian as
\begin{equation}
G(x_1,x_2;z) = \frac{2m}{W(z)}\left[ u_L(x_1;z)u_R(x_2;z)\theta(x_2-x_1)
           + u_L(x_2;z)u_R(x_1;z)\theta(x_1-x_2) \right]\,,
\end{equation}
where $\theta(x)$ is the Heaviside step function.

For the square well considered here, we have \cite{Baltin}
\begin{equation}
\bar{\Delta}_{-\ell/2,\ell/2}
=  -\left(\frac{1}{-\frac{d^2}{dx^2}+m^2+2J_2(x)}\right)_{-\ell/2,\ell/2}
= -\frac{2}{W}\,,
\end{equation}
where
\begin{equation}
W = 4m \left( \cosh (\ell\sqrt{m^2+2mV})
  +  \frac{m^2+mV}{m(\sqrt{m^2+2mV})} \sinh (\ell\sqrt{m^2+2mV}) \right)\,.
\end{equation}
%
For large $\ell$, this expression becomes
\begin{equation}
\bar{\Delta}_{-\ell/2,\ell/2} \approx \frac{-\exp \left( - \ell \sqrt{m^2 + 2 mV} \right) }
{m \left( 1+ \frac{m^2+mV}{m \sqrt{m^2 + 2 mV}} \right)}
\,.
\end{equation}
We can interpret the result by noting that $\bar{\Delta}_{-l/2,l/2}$ scales 
like $\exp\left(-m_{\rm eff} \ell\right)$, where the effective mass of the 
exchanged particle is 
\begin{equation}
m_{\rm eff} = \sqrt{m^2 + 2 mV}\,.
\end{equation}

As well as being suppressed by an additional factor of $\lambda$, this 
contribution to the interaction energy falls off more rapidly with $\ell$ 
than the contribution from the contact term. In $\lambda \phi^4$ theory, 
two particles are exchanged at order $\lambda^2$. 
(see Fig.~\ref{fig:feynman-potential}$b$). Therefore, the interaction energy 
is suppressed by a further factor of $\exp\left(-m_{\rm eff} \ell\right)$.

\subsection{Gate Universality}\label{Subsec:Gate_Universality}

As we saw in the previous subsection, if we attempt to perform operations 
on two qubits by bringing one well from each qubit close
together, the particle is more likely to tunnel than to interact. If the
particle tunnels, then the state will leave the computational
subspace. However, as we show in this subsection, by rotating through a
larger space we can implement a universal gate set on the
computational subspace.
Specifically, we show how one can perform an arbitrary two-qubit gate,
corresponding to a $4 \times 4$ unitary matrix, to polynomial precision by
smoothly varying the well depths and separations. 

As we wish to show
BQP-hardness of scattering, in our BQP-hardness construction we
do not allow access to measurements during the execution of the quantum
circuit. (Allowing measurements would change the BQP-hardness question
substantially. By the KLM construction \cite{KLM01}, one can achieve
computational universality using adaptive measurements on a free field
theory.) Furthermore, we do not assume any reservoir of
cleanly prepared ancilla qubits. Thus we cannot simply invoke
fault-tolerance threshold theorems such as that in \cite{Aharonov_Benor}. 
In the absence of error correction, it suffices to perform each gate in a
quantum circuit of $G$ gates with $O(1/G)$ infidelity
(see, for example, \cite{Nielsen_Chuang}).

For a pair of double wells containing two particles, there are six
nearly degenerate states corresponding to the $\binom{4}{2}$ combinations 
of two identical particles in four wells without double occupation. 
Let us call the subspace spanned by these six states $\mathcal{S}$. 
States involving double occupation of a well are not degenerate with those in
$\mathcal{S}$, since their energy is altered by the interparticle potential
induced by the $\phi^4$ term. Excited bound states of the wells, 
states in which the particles are unbound from the wells, and states 
involving additional particles all have higher energy than those in 
$\mathcal{S}$. 
Thus, by varying the depths and separations of the wells slowly enough 
we can keep the system in the adiabatic regime, with transitions out of 
$\mathcal{S}$ suppressed by the energy gaps separating $\mathcal{S}$ 
from the rest of the spectrum above and below.

The adiabatic theorem proven in \cite{N93} shows that the probability
of leakage out of the subspace $\mathcal{S}$ can be made exponentially
small as a function of how slowly we vary the Hamiltonian, as long as 
the time variation of the Hamiltonian is sufficiently smooth.\footnote{See
also \cite{ASY87,EH12}.} Let $H(s)$ be a parameterized family of
Hamiltonians, and consider the time evolution induced by the
time-dependent Hamiltonian $H(t/\tau)$. If $H(s)$ belongs to the
Gevrey class of order $\alpha$, then the diabatic error scales as
\begin{equation}
\epsilon \sim \exp \left[ - \tau^{\frac{1}{1+\alpha}} \right].
\end{equation}
A function $g(s)$ in the Gevrey class of order $\alpha$ is  a
smooth function on $\mathbb{R}$ such that on any interval $I = [a,b]
\subset \mathbb{R}$ there are constants $C$ and $R$ for which
\begin{equation}
\left| \frac{d^k g}{ds^k} \right| \leq C R^k k^{\alpha k},
\quad \textrm{for $k=1,2,3,\ldots$}
\end{equation}
For $\alpha = 1$ this is the class of analytic functions. For larger
$\alpha$ the condition is less restrictive. In particular, for $\alpha
> 1$ there exist smooth, compactly supported ``bump'' functions 
not identically equal to zero. By varying the well
depths and well separations according to such bump functions, we can
limit leakage errors out of $\mathcal{S}$ to $\epsilon$ at a cost of
$\tau = \mathrm{poly}(\log 1/\epsilon)$. 
Thus, the requirement $\epsilon \sim 1/G$ contributes only a polylogarithmic 
factor to the time needed to execute $G$ gates.

Given that the total amplitude to be outside of the subspace
$\mathcal{S}$ is limited to $\epsilon \sim 1/G$, we can neglect
this amplitude and solve for the dynamics within
$\mathcal{S}$ 
with the approximation 
$\ket{\psi(t)} \in \mathcal{S}(t)$ for all $ t$.
We do so in the adiabatic frame, that is, the instantaneous
eigenbasis of $H(t)$, writing $\mathcal{S} = \mathrm{span} \{
\ket{L_1(t)}, \ldots, \ket{L_6(t)} \}$, with $H(t) \ket{L_j(t)} =
E_j(t) \ket{L_j(t)}$. 
In Appendix~\ref{effschro},
we obtain the following effective
Schr\"{o}dinger equation for the dynamics within $\mathcal{S}(t)$: 
\begin{equation}
\frac{d \ket{\psi}}{dt} = -i H_A(t) \ket{\psi} + O(1/G)\,,
\end{equation}

\begin{equation}
\label{HA}
\bra{L_j(t)} H_A(t) \ket{L_k(t)} = \left\{ \begin{array}{cl}
E_j(t)\,, & \textrm{if $j=k$}, \vspace{5pt} \\
i \frac{\bra{L_j(t)} \frac{dH}{dt} \ket{L_k(t)}}{E_j(t) - E_k(t)}\,, &
\textrm{otherwise}. \end{array} \right.
\end{equation}  

We now consider in more detail how one can implement a universal set of
quantum gates using these dynamics. First, consider single-qubit gates. 
Recall that, if we choose dual-rail encoding, 
logical zero and one are encoded by occupation of the
ground state of the left and right wells, respectively. 
The ground and first excited eigenstates of the double-well Hamiltonian are 
to a very good approximation the symmetric and antisymmetric linear 
combinations of these states, respectively. Their energy separation is 
exponentially small as a function of the separation of the wells. Thus, 
the left and right occupied states are exponentially long-lived and form a
convenient basis in which to work.

\begin{figure}
\begin{center}
\includegraphics[width=0.65\textwidth]{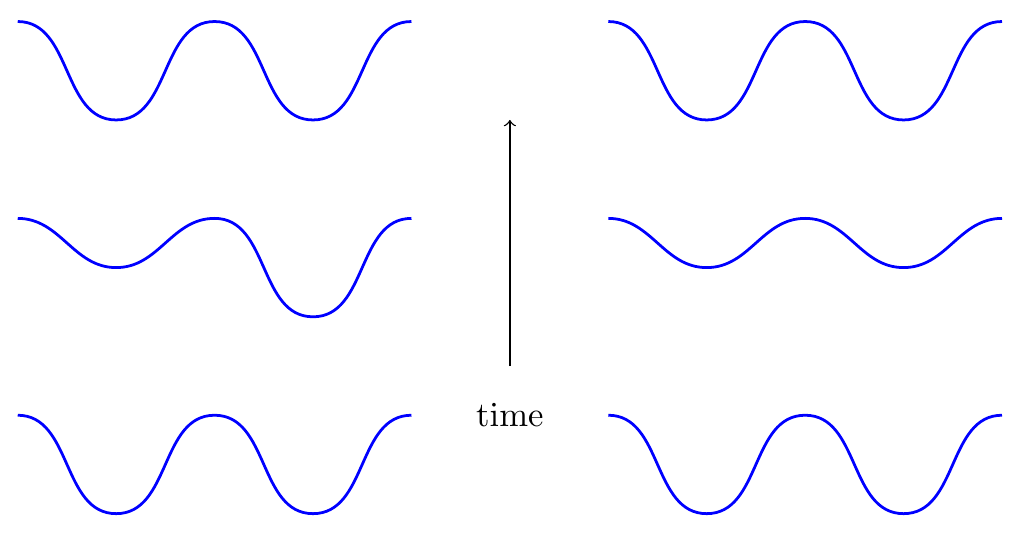}
\caption{\label{figXZ} 
  The variation of well depths implements a $Z$ rotation (left-hand side),
  while lowering the barrier height implements an $X$ rotation (right-hand side). } 
\end{center}
\end{figure}

One can perform logical $Z$ rotations\footnote{Following quantum
  information conventions, we use $X,Y,Z$ rather than $\sigma_x,
  \sigma_y, \sigma_z$ to denote the Pauli matrices.} on a qubit by varying the
depths of the wells (see Fig.~\ref{figXZ}). This
procedure implements 
the rotation $e^{-i Z \theta}$ 
in the logical
basis such that the angle $\theta$ is proportional to the product of
the depth and duration of the variation of the wells. 
To prevent the accumulation of error, one is required by the adiabatic 
theorem to perform the variation of the wells logarithmically
slower as the number of gates in the circuit is increased. To achieve
a fixed target rotation angle, one can decrease the depth of the well 
variation by the same factor by which the duration of the variation is
increased. For a concrete realization of a $Z$ gate in this scheme,
see Appendix~\ref{zgate}.

On can implement an $X$ rotation by temporarily lowering the barrier
between wells (see Fig.~\ref{figXZ}). From \eq{HA} one
sees that, in the limit where the two wells are completely isolated
from all others, the off-diagonal elements of $H_A$ in the eigenbasis
are zero because $dH/dt$ and the ground state have exact
left-right symmetry, whereas the first excited state is
antisymmetric. Thus, $H_A$ implements a pure $Z$ rotation in the
eigenbasis. This corresponds to a pure $X$ rotation in the logical
basis, as this is related to the eigenbasis by the Hadamard
transform 
\begin{equation}
H = \frac{1}{\sqrt{2}} \left[ \begin{array}{rr} 1 & 1 \\ 1 &
    -1 \end{array} \right].
\end{equation} 
One can accommodate the limits on gate speed imposed by adiabaticity 
while still implementing the desired rotation angle by
correspondingly adjusting the degree of lowering of the barrier. For a
concrete quantitative realization of an $X$ gate in this manner, see
Appendix~\ref{xgate}.

\begin{figure}
\begin{center}
\includegraphics[width=0.65\textwidth]{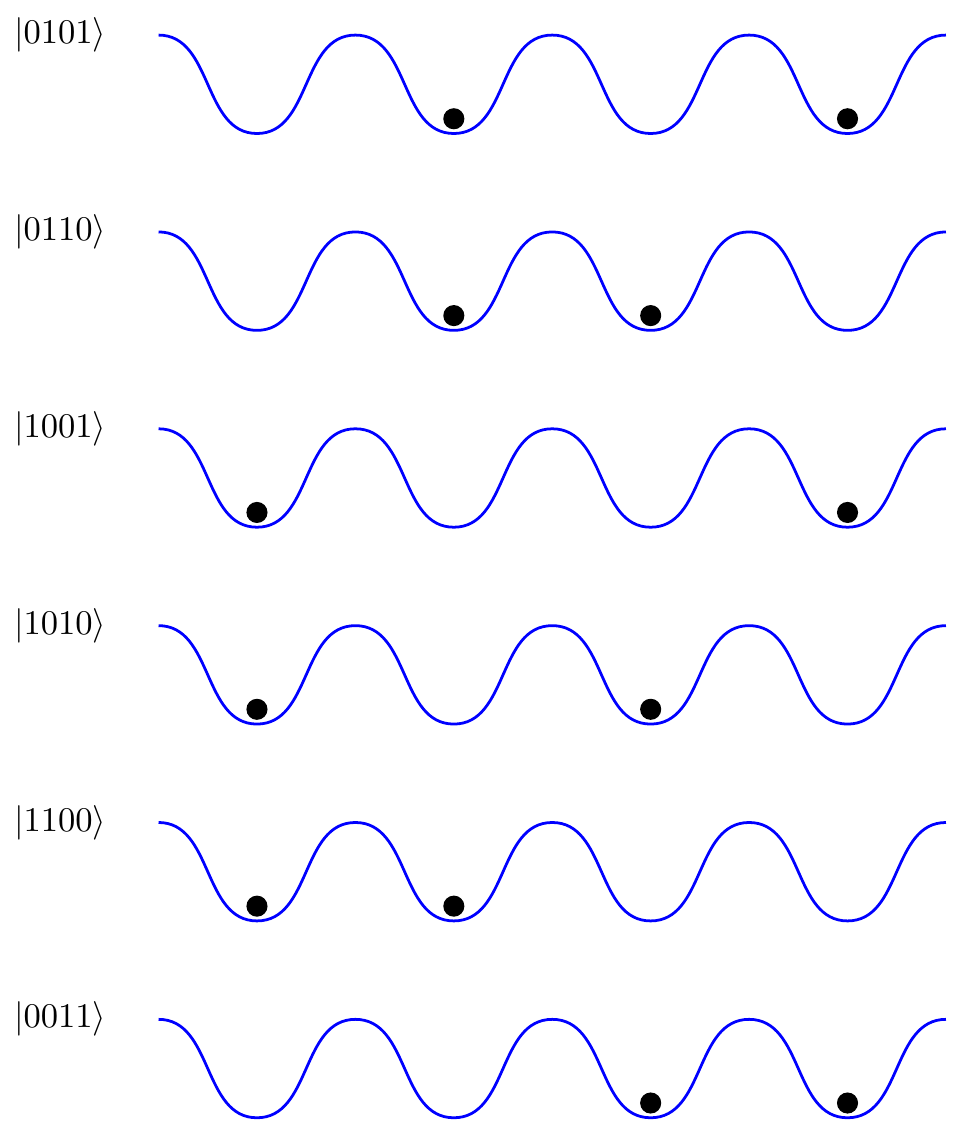}
\caption{\label{ocbase} The occupation-number basis for the
  four wells associated with a pair of logical qubits. 
  In dual-rail encoding,
  the occupation-number states
  $\ket{0101}$, $\ket{0110}$, $\ket{1001}$, and $\ket{1010}$ encode
  the logical qubit states $\ket{11}$, $\ket{10}$, $\ket{01}$, and
  $\ket{00}$, respectively. 
  The occupation-number states $\ket{1100}$ and $\ket{0011}$ lie outside 
  the coding subspace but are unavoidable because of tunnelling.} 
\end{center}
\end{figure}

From $X$ and $Z$ rotations, one can construct an arbitrary single-qubit
gate using Euler angles. Any entangling two-qubit gate yields a
universal quantum gate set when supplemented by arbitrary single-qubit
gates \cite{BB02}. Thus the final task of this subsection is to construct
an entangling two-qubit gate through the use of the inter-particle
interaction. We perform the analysis in the
occupation-number basis for $\mathcal{S}$ (Fig.~\ref{ocbase}).


\begin{figure}
\begin{center}
\includegraphics[width=0.65\textwidth]{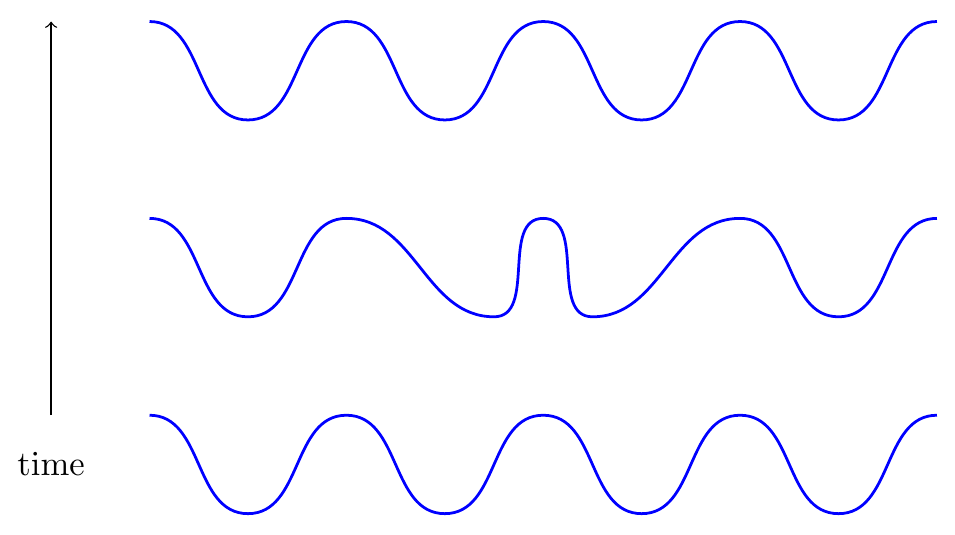}
\caption{\label{cphase} 
Our implementation of
an entangling two-qubit gate. If the center two wells are occupied
(corresponding to the logical $\ket{10}$ state), 
the attraction between particles induces a phase rotation. 
In the case that exactly
one of the two center wells is occupied, there is a tunneling amplitude
into the noncoding subspace.} 
\end{center}
\end{figure}

We perform a two-qubit gate by temporarily decreasing the separation
between the two center wells in the quadruple-well system
(Fig.~\ref{cphase}). The induced Hamiltonian
on $\mathcal{S}$ in the occupation-number basis takes the form
\begin{equation} \label{HAmat}
H_A(t) \simeq \begin{array}{rl} \left[ \begin{array}{cccccc}
  0 & 0 & 0 & 0 & 0 & b(t) \\
  0 & c(t) & 0 & 0 & 0 & 0 \\
  0 & 0 & d(t) & 0 & 0 & 0 \\
  0 & 0 & 0 & 0 & b & 0 \\
  0 & 0 & 0 & b(t) & 0 & 0 \\
  b(t) & 0 & 0 & 0 & 0 & 0
\end{array} \right] & \begin{array}{l}
0101 \\
0110 \\
1001 \\
1010 \\
1100 \\
0011
\end{array} \end{array}
\end{equation}
for some time-dependent coefficients $b(t)$, $c(t)$ and $d(t)$, which 
depend on the choice of the shapes of the wells and their trajectories
and can be determined numerically from \eq{HA}. 
Here, exponentially suppressed tunneling matrix elements between distant wells 
have been neglected.
The off-diagonal $b(t)$ entries describe the tunneling transition 
$|01\rangle\leftrightarrow|10\rangle$ for the two wells that approach each 
other. 
In addition, there are on-diagonal contributions because the energy changes 
slightly as the wells get closer together.  We have defined our 
(time-dependent) zero of energy so that this energy shift vanishes when one 
of the two wells is occupied and the other is empty. Therefore, the only 
nonzero diagonal entries in $H_A(t)$ are those where both wells are occupied
or both are empty, denoted $c(t)$ and $d(t)$, respectively. 

Recall that we encode a qubit by placing a single particle in either one 
of two adjacent potential wells. Thus, the two-qubit Hilbert space is spanned 
by the four states $\{|0101\rangle,|0110\rangle,|1001\rangle,|1010\rangle\}$, 
while the two states $\{|1100\rangle,|0011\rangle\}$ are not valid encodings. 
We want to execute an entangling gate that preserves the valid two-qubit 
subspace. We note that the unitary time evolution induced by the 
time-dependent Hamiltonian $H_A(t)$ is a direct sum of a 
diagonal transformation acting on $\{ \ket{0110}, \ket{1001} \}$
and two identical
$X$ rotations $e^{i X \theta_x}$ in $\mathrm{span} \{ \ket{0101},
\ket{0011} \}$ and $\mathrm{span} \{ \ket{1010}, \ket{1100} \}$. 
If we replace $H_A(t)$ with $H_A(t/z)$, thus increasing the duration of the 
execution of the gate by the factor $z$, then the rotation angle $\theta_x$ 
also increases by the factor $z$. We choose $z$ sufficiently large that our 
adiabaticity constraint is satisfied and tune its value so that 
$\theta_x/2 \pi$ is an integer. The resulting unitary transformation $U$ 
preserves the four-dimensional subspace spanned by our two encoded qubits; 
acting on 
$\mathrm{span}\{|0101\rangle,|0110\rangle,|1001\rangle,|1010\rangle\}$, 
it is the diagonal gate
\begin{equation}
U = {\rm diag}\left(1, e^{i\alpha},e^{i\beta},1\right),
\end{equation}
where we evaluate the phases $\alpha$ and $\beta$ by integrating $c(t)$ and 
$d(t)$, respectively. This $U$ is an entangling gate unless 
$e^{i(\alpha + \beta)}=1$, which will not be satisfied for a generic choice 
of the shapes and trajectories of the wells.

Because the interaction strength is $O(\lambda)$, the time taken 
to execute a single entangling gate with an $O(1)$ phase shift is
at least $O(1/\lambda)$.
Adiabatic protection against leakage from the coding subspace imposes 
a stronger lower bound on the gate duration. The energy gap $\gamma$ 
separating the doubly occupied states is of order $\lambda$, and hence 
the runtime must scale as
$\widetilde{O}(\gamma^{-2}) = \widetilde{O}(\lambda^{-2})$. In a
circuit with $G$ gates, we need to choose $\lambda = O(1/G)$ to
justify neglecting corrections that are higher order in $\lambda$ when
computing the form of the source term $J_2$ needed to implement a
given gate with infidelity $O(1/G)$. 
Therefore, the simulation time is $O(G^2)$ for a single gate and 
$O(G^2D)$ for the complete circuit, where $D$ is the circuit depth 
(not including the state-preparation step analyzed in 
Sec.~\ref{sec:interact}).  

Furthermore, we note that our two-qubit gates are geometrically local in 
one dimension: only neighboring qubits interact. To perform our entangling 
two-qubit gate on two distantly separated qubits $A$ and $B$, we would 
perform a series of swap gates to bring $A$ and $B$ into neighboring 
positions, execute the entangling gate, and then use swap gates to return 
$A$ and $B$ to their original positions. The swap gate can be well 
approximated via our universal gate repertoire with only polylogarithmic 
overhead, but the swaps would increase the circuit depth by a factor $O(n)$, 
where $n$ is the number of qubits, compared with a circuit with nonlocal 
two-qubit gates. 

\section{Measurements and BQP-completeness}\label{sec:measurement}

The main line of reasoning in this paper establishes a BQP-hardness
argument for the problem of determining transition probabilities in 
$(1+1)$-dimensional $\phi^4$ theory to polynomial precision. To obtain a
BQP-completeness result we need to establish that (the decision version of)
this problem is also contained in BQP, that is, that it can be solved by a
polynomial-time quantum algorithm. Essentially, the quantum algorithm
achieving this is given in \cite{JLP12, JLP14}. However, there are
some small differences between the BQP-hard problem given in 
this paper and the problem solved by the
quantum algorithm of \cite{JLP12, JLP14}. Here, we address several
ways to bridge this gap, that is, several problems that one can show
to be BQP-complete by technical variations of the algorithms of
\cite{JLP12, JLP14} and the argument of the preceding sections.

In our main BQP-hardness argument, given a quantum circuit, we
construct $J_1(t,x)$ and $J_2(t,x)$ so that the vacuum-to-vacuum
transition amplitude approximates the $\ket{0\ldots0}$ to
$\ket{0\ldots0}$ amplitude of the quantum circuit. 
The problem of deciding whether the magnitude of the amplitude 
$\ket{0\ldots0} \to \ket{0\ldots0}$
of a quantum circuit is smaller than 1/3 or larger than
2/3, given the promise that one of these is the case, is
BQP-complete. As an immediate corollary, it is BQP-hard to 
estimate the corresponding vacuum-to-vacuum transition probability
to within  $\pm \epsilon$, for a sufficiently small constant $\epsilon$.

\begin{figure}
\begin{center}
\includegraphics[width=0.5\textwidth]{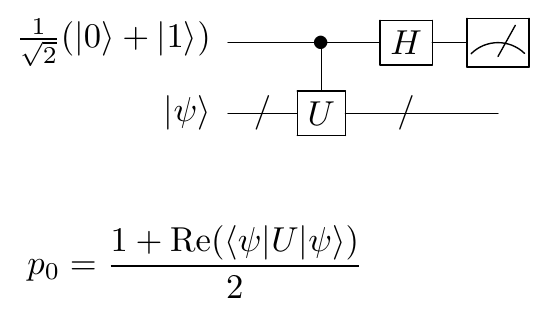}
\caption{\label{hadtest} Circuit implementing the Hadamard test. 
The boxed $H$ and $U$ denote Hadamard and controlled-$U$ gates, and the 
slashed line indicates multiple qubits. 
The probability of measuring $\ket{0}$ is denoted $p_0$.
}
\end{center}
\end{figure}

The existence of a polynomial-time quantum algorithm that estimates the 
vacuum-to-vacuum transition amplitude to adequate precision would
imply that the 
decision problem is not only BQP-hard but also BQP-complete. 
One can devise such an algorithm using the methods of 
\cite{JLP12, JLP14}, 
which described quantum algorithms 
for preparing the vacuum state and implementing the unitary time evolution 
in $\phi^4$ theory. 
This procedure can be applied without modification in the 
presence of the spacetime-dependent source terms $J_1(t,x)$ and $J_2(t,x)$, 
at the cost of modest performance penalties, by the analysis of 
\cite{suzuki:1993, wiebe:2010}. 
With these tools from \cite{JLP12, JLP14}, 
we can construct the algorithm for estimating the 
amplitude $\langle {\rm vac}|U|{\rm vac}\rangle$ using a standard technique, 
called the {\it Hadamard test}, which is illustrated in 
Fig.~\ref{hadtest}. 
The probability of measuring $\ket{0}$ is 
$p_0= [1+\mathrm{Re}(\bra{\psi}U\ket{\psi})]/2$.
Thus, one can obtain the real part of
$\bra{\psi}U\ket{\psi}$ to within $\epsilon$ by making
$O(1/\epsilon^2)$ measurements. Similarly, by initializing the
control qubit to $\frac{1}{\sqrt{2}} (\ket{0} - i \ket{1})$,
one can estimate the imaginary part of $\bra{\psi}U\ket{\psi}$.

If we can prepare the state $|\psi\rangle$, 
execute the conditional unitary transformation $U$ controlled by a single 
qubit, namely,
\begin{equation}
{\rm controlled-}U = |0\rangle\langle 0| \otimes I 
		+ |1\rangle \langle 1| \otimes U,
\end{equation}
and measure the control qubit, then we can estimate 
$\langle \psi|U|\psi\rangle$ using the Hadamard test. 
We can promote the circuit described in \cite{JLP12, JLP14} 
for implementing the unitary time-evolution operator $U$ 
to a circuit for controlled-$U$ by replacing each gate $G$ in the circuit 
with controlled-$G$. (If $G$ is a two-qubit gate, then controlled-$G$ is 
a three-qubit gate, which can be efficiently decomposed into the original 
gate set through standard techniques.) Therefore, the methods described 
in \cite{JLP12, JLP14} 
for preparing the vacuum and for implementing time evolution, 
together with the Hadamard test, provide a procedure for estimating the 
vacuum-to-vacuum amplitude in the presence of sources that vary in space 
and time. This procedure, combined with the result obtained in this paper 
that the problem is BQP-hard, shows that the corresponding decision problem
is BQP-complete. 

This scheme for demonstrating the BQP-completeness of a quantum field theory 
problem has the advantage that only a single-qubit measurement 
is required to read out the result, but also the disadvantage that each gate 
needs to be replaced by its controlled version. There are other ways to 
bridge the gap between the BQP-hardness result presented in this paper and 
the simulation algorithms formulated in \cite{JLP12, JLP14}, 
in which we avoid the nuisance 
of replacing each $G$ by controlled-$G$ at the cost of executing a more 
complicated measurement at the end of the algorithm. For example, we could 
omit the final step of the BQP-hardness construction, in which particles in 
the logical-zero states are annihilated through adiabatic passage. In 
that case, the transition probability that is BQP-hard to estimate is the 
probability to start in the vacuum and end with all of the 
double wells in the logical-zero state.
The algorithm that estimates this transition probability 
includes a final step that simulates, through phase estimation, a particle 
detector measuring the energy in a spatially localized region. This detector 
simulation was described in \cite{JLP14}. 
Another motivation 
for discussing particle-detecting measurements is that our BQP-hardness 
construction can be regarded as an idealized architecture for
constructing a universal quantum computer from laboratory systems, 
namely, condensed-matter or atomic-physics experimental 
platforms that are described by $\phi^4$ theory. With these 
motivations in mind, we briefly explain, following \cite{JLP14}, 
how to measure the energy in a local region.

At the end of our scattering process, the Hamiltonian is
\begin{equation}
H = \int dx \mathcal{H}(x) 
\,,
\end{equation}
where
\begin{equation}
\mathcal{H}(x) = \frac{1}{2} \pi^2(x) + \frac{1}{2} \phi(x)
  \frac{d^2}{dx^2} \phi(x) + \frac{1}{2} m^2 \phi^2(x)
 + J_2(x) \phi^2(x) + \frac{\lambda}{4!} \phi^4(x).
\end{equation}
Measuring the observable $H$ would yield the total energy of the
system. Correspondingly, consider the operator
\begin{equation}
H_f = \int dx f(x) \mathcal{H}(x) 
\,,
\end{equation}
where $f$ is some envelope function that is 
localized in some spatial region $R$.
Then, measuring the observable $H_f$ yields an approximation to
the energy within $R$. If a particle is present within $R$, then
the measured energy will be detectably larger than if $H_f$ is
measured for vacuum.

The envelope function $f$ must be chosen with care. The naive choice
\begin{equation}
\label{sharp}
f(x) = \left\{ \begin{array}{rl} 1, & x \in R, \\
0, & \textrm{otherwise}
\end{array} \right.
\end{equation}
is unsuitable because of the $\phi(x) \frac{d^2}{dx^2} \phi(x)$ term and 
the discontinuity in $f$. More quantitatively, one can introduce a lattice
cutoff, as is done in \cite{JLP14}, and compute the variance of $H_f$ in
the vacuum state. For the functional form (\ref{sharp}), this diverges as
the lattice spacing is taken to zero. In one spatial dimension, one can
obtain a convergent variance by instead choosing $f$ to be a
Gaussian envelope.

With the aim of interpreting $H_f$, it is helpful to consider
the free theory ($\lambda = 0$), which can be exactly solved. 
Associated with each
potential well created by $J_2(x)$ there is at least one localized mode
representing a particle bound in this well. The creation and
annihilation operators associated with this mode can be expanded in
the form
\begin{equation}
a^\dag = \int dx \left[ c_{\phi}(x) \phi(x) + c_{\pi}(x) \pi(x) \right].
\end{equation}
The magnitudes of $c_{\phi}$ and $c_{\pi}$ decay exponentially with
characteristic decay length $1/m$ outside the well. Hence, if the
Gaussian envelope is centered on the well and has a large width relative
to the width of the well plus $1/m$, then $[a^\dag,H_f] \simeq [a^\dag, H]$ 
and $[a,H_f] \simeq [a,H]$. This in turn implies that the presence of a
particle in the well raises the expectation value of $H_f$ by
approximately the same amount that the expectation of $H$ is raised,
that is, by the energy of the particle. In the $\lambda \neq 0$ case,
the qualitative behavior will be similar.\\

\bigskip
\noindent
\textbf{Acknowledgments:} 
We thank David Gosset, Mark Rudner, and Jacob Taylor for helpful discussions.
JP gratefully acknowledges support from the Institute for Quantum 
Information and Matter (IQIM), an NSF Physics Frontiers Center with support 
from the Gordon and Betty Moore Foundation, from the Army Research Office, 
and from the Simons Foundation It from Qubit Collaboration.
KL was supported in part by NSERC and the Centre for Quantum Information
and Quantum Control (CQIQC).
Parts of this manuscript are a contribution of NIST, an agency of the US
government, and are not subject to US copyright.

\appendix

\section{Dynamics within an Adiabatic Subspace}
\label{effschro}

Suppose that we have a time-dependent Hamiltonian in which some subset
of eigen-energies are at all times well separated from the rest of the
spectrum. Then, using adiabatic theorems, one can prove that the
amplitude to escape the isolated subspace is always less than
some bound $\epsilon$. In this appendix, we adapt fairly standard
arguments to prove that, in this circumstance, the dynamics within the
isolated subspace induced by an adiabatic process of duration $t$ is
given by \eq{HA}, up to errors of order $\epsilon t$.

Let $H(t)$ be a differentiable time-dependent Hamiltonian. Let
$\ket{L_1(t)}$, $\ket{L_2(t)}$, $\ldots$ be a normalized eigenbasis
with corresponding eigenvalues $E_1(t)$, $E_2(t)$, $\ldots$, so that,
for all $j$ and all $t$,
\begin{equation}
\label{eigen}
H(t) \ket{L_j(t)} = E_j(t) \ket{L_j(t)} \,,
\end{equation}
with
\begin{equation}
\label{norm}
\langle L_j(t) | L_j(t) \rangle = 1.
\end{equation}
The normalization condition \eq{norm} leaves us free to choose the
phase. If $\ket{L_k'(t)}$ is a normalized eigenvector of $H(t)$, then
so is 
\begin{equation}
\label{sois}
\ket{L_k(t)} = e^{i \vartheta(t)} \ket{L_k'(t)}
\end{equation}
for any real-valued function $\vartheta(t)$. Henceforth, we shall generally
leave the time dependence implicit in $\ket{L_k'}$.  
Differentiating the normalization condition $\langle L_k' |
L_k' \rangle = 1$ yields
\begin{equation}
\frac{d \bra{L_k'}}{dt} \ket{L_k'} + \bra{L_k'} \frac{d
  \ket{L_k'}}{dt} = 0.
\end{equation}
Thus, for all $k$ and $t$,
\begin{equation}
\label{purim}
\mathrm{Re} \left[ \bra{L_k'} \frac{d \ket{L_k'}}{dt} \right] = 0.
\end{equation}
By \eq{purim}, we can let
\begin{equation}
\vartheta(t) = i \int_0^t ds \bra{L_k'(s)} \frac{d \ket{L_k'(s)}}{ds} \,,
\end{equation}
which is purely real and hence gives a normalized $\ket{L_k(t)}$ via
\eq{sois}. This choice of $\vartheta(t)$ yields
\begin{eqnarray}
\bra{L_k(t)} \frac{d \ket{L_k}}{dt} & = & \bra{L_k'} e^{-i
  \vartheta(t)} \frac{d}{dt} e^{i \vartheta(t)} \ket{L_k'} \\
& = & i \frac{d \vartheta}{dt} + \bra{L_k'} \frac{ d
  \ket{L_k'}}{dt} \\
& = & 0.
\end{eqnarray}
Thus we are free to choose
\begin{equation}
\label{phasecon}
\bra{L_k} \frac{d \ket{L_k}}{dt} = 0\,, \quad \forall k,t,
\end{equation}
as this is ultimately a phase convention. We shall adopt this
convention for the remainder of this section.

Next, following standard treatments (see, for example, \cite{Messiah}) we
express the dynamics in the instantaneous
eigenbasis\footnote{We shall refer to this as the adiabatic frame. Some older 
references such as \cite{Messiah} refer to this as the rotating-axis 
representation.} of $H(t)$, that is,  
\begin{equation}
\label{adframe}
\ket{\psi(t)} = \sum_j \alpha_j(t) \ket{L_j(t)}.
\end{equation}
By \eq{adframe} and Schr{\"o}dinger's equation,
\begin{eqnarray}
\frac{d \alpha_j}{dt} & = & \frac{d}{dt} \langle L_j | \psi \rangle \\
& = & \frac{d \bra{L_j}}{dt} \ket{\psi} - i \bra{L_j} H \ket{\psi}.
\end{eqnarray}
By Eqs.~(\ref{eigen}) and (\ref{adframe}), this simplifies to
\begin{equation}
\label{alphadot}
\frac{d \alpha_j}{dt} = \frac{d \bra{L_j}}{dt} \ket{\psi} - i E_j \alpha_j.
\end{equation}

Differentiating \eq{eigen} yields
\begin{equation}
\frac{dH}{dt} \ket{L_j} + H \frac{d \ket{L_j}}{dt} = \frac{dE}{dt}
  \ket{L_j} + E_j \frac{d \ket{L_j}}{dt}.
\end{equation}
Thus,
\begin{equation}
\label{so}
(H-E_j) \frac{ d \ket{L_j}}{dt} = \left( \frac{dE_j}{dt} -
  \frac{dH}{dt} \right) \ket{L_j}.
\end{equation}
Let
\begin{equation}
\label{Gdef}
G_j = \sum_{k \neq j} \frac{P_k}{E_k-E_j}\,,
\end{equation}
where $P_k = \ket{L_k} \bra{L_k}$ is the projector on to the $k\th$
eigenspace. Then
\begin{equation}
G_j (H-E_j) = \id - P_j\,.
\end{equation}
Thus, multiplying \eq{so} by $G_j$ yields
\begin{equation}
(\id - P_j) \frac{d \ket{L_j}}{dt} = G_j \left( \frac{d E_j}{dt} -
  \frac{dH}{dt} \right) \ket{L_j}\,.
\end{equation}
By our phase convention \eq{phasecon}, this simplifies to
\begin{equation}
\frac{d \ket{L_j}}{dt} = G_j \left( \frac{d E_j}{dt} - \frac{dH}{dt}
\right) \ket{L_j}.
\end{equation}
The first term on the right-hand side vanishes because $G_j$ commutes
with $dE_j/dt$ 
(which is just a c-number) and projects out
$\ket{L_j}$. Thus
\begin{equation}
\label{Ldot}
\frac{d \ket{L_j}}{dt} = - G_j \frac{dH}{dt} \ket{L_j}.
\end{equation}
Substituting \eq{Ldot} into \eq{alphadot} yields
\begin{equation}
\label{alphadot2}
\frac{d \alpha_j}{dt} = - \bra{L_j} \frac{dH}{dt} G_j \ket{\psi} - i
\alpha_j E_j.
\end{equation}
By Eqs.~(\ref{adframe}) and (\ref{Gdef}),
\begin{equation}
G_j \ket{\psi} = \sum_{k \neq j} \frac{\alpha_k}{E_k - E_j} \ket{L_k},
\end{equation}
Therefore \eq{alphadot2} yields
\begin{equation}
\label{finalphadot}
\frac{d \alpha_j}{dt} = -i \alpha_j E_j - \sum_{k \neq j}
\frac{\bra{L_j} \frac{dH}{dt} \ket{L_k}}{E_k -E_j} \alpha_k.
\end{equation}
We thus have a Schr{\"o}dinger-like equation for the coefficients
$\alpha_1,\alpha_2,\ldots$, namely,
\begin{equation}
\label{Mschrod}
\frac{d}{dt} \left[ \begin{array}{c} \alpha_1 \\ \alpha_2 \\
    \vdots \end{array} \right] = - i M \left[ \begin{array}{c} \alpha_1 
    \\ \alpha_2 \\ \vdots \end{array} \right] \,,
\end{equation}
where $M$ is the Hermitian matrix
\begin{equation}
\label{Melements}
M_{jk} = \left\{ \begin{array}{cl}
E_j, & \textrm{if $j=k$,} \vspace{3pt} \\
i \frac{ \bra{L_j} \frac{dH}{dt} \ket{L_k}}{E_j-E_k} \,, & 
\textrm{if $j \neq k$}.
\end{array} \right.
\end{equation}

So far, our analysis has been only a change of frame, maintaining both
exactness and a high degree of generality.\footnote{We have not
  maintained full generality in 
  Eqs.~(\ref{Mschrod}) and (\ref{Melements}) 
  in that it has been convenient to make the technical assumption that
  the spectrum of $H(t)$ is fully discrete and nondegenerate. This is
  generally not true of quantum field theories, but it suffices for
  our analysis in Sec.~\ref{Subsec:Gate_Universality} because the
  relevant low-lying spectrum is discrete and nondegenerate.} Now,
assume that the eigenstates $\ket{L_1(t)}, \ldots, \ket{L_d(t)}$ are
separated from the rest of the spectrum by an energy gap
sufficiently large that, by an adiabatic theorem (for example, \cite{N93}), 
a state initially within the span of $\ket{L_1(t)}, \ldots,
\ket{L_d(t)}$ remains within this subspace for all time, up to
corrections of order $\epsilon$. The dynamics in the adiabatic
frame is then of the form
\begin{equation}
\frac{d}{dt} \left[ \begin{array}{c} \alpha_- \\ \alpha_+ \end{array} \right]
 = -i \left[ \begin{array}{cc} M_{--} & M_{-+} \\ M_{+-} & M_{++} \end{array} 
\right]
\left[ \begin{array}{c} \alpha_- \\ \alpha_+ \end{array} \right]
\,,
\end{equation}
where $M_{--}$ is a $d \times d$ matrix, $\alpha_-$ is a
$d$-dimensional column vector, etc. 

Next, we make an approximation: we suppose that $\alpha_+$ can be
neglected, because the adiabatic theorem guarantees that the magnitude
of the vector $\ket{\alpha_+} = \sum_{j>d} \alpha_j \ket{L_j(t)}$ is at
most $\epsilon$. In other words, we approximate the full dynamics on
$\mathcal{H}$ by a self-contained dynamics within $\mathcal{H}_-(t)$,
namely,
\begin{equation}
\ket{\bar{\psi}} = \sum_{j=1}^d \bar{\alpha}(t) \ket{L_j}\,,
\end{equation}
where $\bar{\alpha}_j(0) = \alpha_j(0)$ and
\begin{equation}
\label{minuschrod}
\frac{d}{dt} \ket{\bar{\psi}} = -i M_{--} \ket{\bar{\psi}}.
\end{equation}
Thus $\ket{\bar{\psi}(t)}$ is an approximation to the exact state
$\ket{\psi(t)}$ with corrections of order $\epsilon$. More
quantitatively, if $\ket{\psi(0)} \in \mathcal{H}_-$
then
\begin{equation}
\label{initprod}
\langle \bar{\psi}(0) | \psi(0) \rangle = 1
\end{equation}
and
\begin{eqnarray}
\frac{d}{dt} \langle \bar{\psi} | \psi \rangle & = & 
\frac{d \bra{\bar{\psi}}}{dt} \ket{\psi} + \bra{\bar{\psi}} \frac{d \ket{\psi}}{dt} \\
& = & i \bra{\bar{\psi}} M_{--}^\dag \ket{\psi} - i \bra{\bar{\psi}} \left(
  M_{--} \ket{\psi_-} + M_{-+} \ket{\psi_+} \right) \\
& = & -i \bra{\bar{\psi}} M_{-+} \ket{\psi_+},
\end{eqnarray}
where we have used \eq{minuschrod} and the fact that $M_{--}$ is
Hermitian. Combined with \eq{initprod}, this yields
\begin{equation}
\langle \bar{\psi}(t) | \psi(t) \rangle = 1 - i \int_0^t d \tau
\bra{\bar{\psi}} M_{-+} \ket{\psi_+}\,.
\end{equation}
Thus,
\begin{equation}
\label{totalbound}
| \langle \bar{\psi}(t) | \psi(t) \rangle | \geq 1 - \int_0^t \|
M_{-+} \ket{\psi_+} \|\,.
\end{equation}
By assumption, an adiabatic theorem tells us that $\| \ket{\psi_+} \|
\leq \epsilon$ for all $t$, and therefore the above inequality 
implies
\begin{equation}
| \langle \bar{\psi}(t) | \psi(t) \rangle | = 1- O(\epsilon t).
\end{equation}
If $M_{-+}$ is a bounded operator, then \eq{totalbound} yields
\begin{equation}
\label{Mbound}
| \langle \bar{\psi}(t) | \psi(t) \rangle | \geq 1- \epsilon t \|
M_{-+} \|.
\end{equation}
Examining \eq{Melements}, one sees that if $dH/dt$ 
is a bounded operator (as is the case if it is finite-dimensional), one
obtains from \eq{Mbound} the following:
\begin{equation}
| \langle \bar{\psi}(t) | \psi(t) \rangle | \geq 1 - \frac{\left\|
  \frac{dH}{dt} \right\|}{\gamma} \epsilon t \,,
\end{equation}
where
\begin{eqnarray}
\gamma(t) & = & E_{d+1}(t) - E_d(t)\,, \\
\gamma & = & \min_t \gamma(t).
\end{eqnarray}

\section{Fourier Transform of the Source}
\label{app:fourier}
Here we verify the properties, stated in Sec.~\ref{sec:interact}, of the 
Fourier transform of the source used in adiabatic passage. 
For convenience, we choose the normalization
\begin{align}
f(t) & = \frac{2}{\sqrt{T}} \text{rect}\Big(\frac{t}{T}\Big)
         \cos(\omega_0 t + \kappa t^2/2) \,, 
\end{align}
for which the integral of $|f(t)|^2$ is O(1), independent of $T$.
The Fourier transform of $f(t)$ is then
\begin{equation}
{\cal F}(\omega) = {\cal G}_+(\omega-\omega_0) + {\cal G}_-(\omega+\omega_0)
\,,
\end{equation}
where ${\cal G}_\pm(\omega)$ is the Fourier transform of
\begin{equation}
g_\pm(t) = \frac{1}{\sqrt{T}} \text{rect}\Big(\frac{t}{T}\Big)
	   e^{\pm i \kappa t^2/2} \,.
\end{equation}
Integration gives
\begin{equation}
{\cal G}_\pm(\omega) = \sqrt{\frac{\pi}{\kappa T}} e^{\mp i \omega^2/2\kappa} 
 \big\{ C(x^\pm) + C(x^\mp) \pm i \left[S(x^\pm) + S(x^\mp)\right] 
 \big\} \,,
\end{equation}
where
\begin{align}
x^\pm &=  \sqrt{\frac{\kappa}{\pi}} \left(\frac{T}{2} 
	\mp \frac{\omega}{\kappa} \right) \,.
\end{align}
The special functions $C(z)$ and $S(z)$ are the Fresnel integrals, defined as
\begin{align}
C(z) &= \int_0^z \cos\Big(\frac{1}{2}\pi t^2\Big) dt \,, \\
S(z) &= \int_0^z \sin\Big(\frac{1}{2}\pi t^2\Big) dt \,.
\end{align}
Thus, with $B=\kappa T$, the magnitude of $G_\pm$ is
\begin{align} \label{eq:Gplus}
\frac{B}{\pi} |{\cal G}_\pm(\omega)|^2 &= 
\Big\{C\Big[\Big({\frac{BT}{\pi}}\Big)^{1/2}
\Big(\frac{1}{2}-\frac{\omega}{B}\Big)\Big] 
+ C\Big[\Big({\frac{BT}{\pi}}\Big)^{1/2}
\Big(\frac{1}{2}+\frac{\omega}{B}\Big)\Big]\Big\}^2
\\
& 
+ \Big\{S\Big[\Big({\frac{BT}{\pi}}\Big)^{1/2}
\Big(\frac{1}{2}-\frac{\omega}{B}\Big)\Big] 
+ S\Big[\Big({\frac{BT}{\pi}}\Big)^{1/2}
\Big(\frac{1}{2}+\frac{\omega}{B}\Big)\Big]\Big\}^2
\,. \nonumber
\end{align}


We now consider the behavior of Eq.~(\ref{eq:Gplus}) for large but
finite values of $T$,
with $B$ fixed. 
For small arguments of the Fresnel integrals, we can use the series
expansions
\begin{align}
C(z) &= \sum_{n=0}^{\infty} 
		\frac{(-1)^n (\pi/2)^{2n}}{(2n)!(4n+1)} z^{4n+1}
\,, \label{eq:Cseries}\\
S(z) &= \sum_{n=0}^{\infty} 
		\frac{(-1)^n (\pi/2)^{2n+1}}{(2n+1)!(4n+3)} z^{4n+3}
\,, \label{eq:Sseries}
\end{align}
which converge for all finite values of $z$.
From Eqs.~(\ref{eq:Cseries}) and (\ref{eq:Sseries}), we see the following: 
if $z$ is real with $z^4< (8/\pi^2)m(2m-1)(4m+1)/(4m-3)$ and 
$z^4< (8/\pi^2)m(2m+1)(4m+3)/(4m-1)$, respectively, then
keeping only the first $m$ terms gives an error with the same sign as
and bounded in magnitude by the $m\th$ term.

For large arguments, we use
\begin{align}
C(z) &= \frac{1}{2} + f(z)\sin\Big(\frac{1}{2}\pi z^2\Big) 
	- g(z)\cos\Big(\frac{1}{2}\pi z^2\Big) \,,
\\
S(z) &= \frac{1}{2} - f(z)\cos\Big(\frac{1}{2}\pi z^2\Big) 
	- g(z)\sin\Big(\frac{1}{2}\pi z^2\Big) \,,
\end{align}
with the asymptotic expansions
\begin{align} \label{eq:f}
f(z) &\sim \frac{1}{\pi z} \sum_{m=0}^{\infty} (-1)^m 
	\frac{\left(\frac{1}{2}\right)_{2m}}{(\pi z^2/2)^{2m}}
\,, \\
g(z) &\sim \frac{1}{\pi z} \sum_{m=0}^{\infty} (-1)^m 
	\frac{\left(\frac{1}{2}\right)_{2m+1}}{(\pi z^2/2)^{2m+1}}
\,,
   \label{eq:g}
\end{align}
as $z \to \infty$.
Here, 
$(\alpha)_0 = 1$ and $(\alpha)_n = \alpha(\alpha+1)(\alpha+2) \cdots
(\alpha+n-1)$, $n=1,2,3, \ldots$.
When $z$ is a positive real number, truncation of Eqs.~(\ref{eq:f}) and
(\ref{eq:g}) gives an error with the same sign as and bounded in magnitude
by the first neglected terms \cite{Olver:2010:NHMF}.
This property of the remainder terms is used below.
%

Now, ${\cal G}_\pm(\omega) = {\cal G}_\pm(-\omega)$, so consider the region
$\omega/B \geq 0$. If $1/2- \omega/B \gg \sqrt{{\pi}/{BT}}$ 
for some large, finite value of $T$,
then
\begin{align}
\frac{B}{2\pi} |{\cal G}_\pm(\omega)|^2 &= \frac{c_{-3}}{(BT)^3} 
+ \frac{c_{-2}}{(BT)^2} 
+ \frac{c_{-3/2}}{(BT)^{3/2}} +  \frac{c_{-1}}{BT}   
+ \frac{c_{-1/2}}{(BT)^{1/2}} + 1 \,, 
  \label{eq:Gmag1}
\end{align}
where the coefficients $c_i=c_i(B,T,\omega)$ are
\begin{align}
  \label{eq:Gmag1a}
c_{-3} &= \frac{64 \eta_2^2}{\pi(1-4\hat{\omega}^2)^6}
\big(1+60\hat{\omega}^2 + 240\hat{\omega}^4 + 64\hat{\omega}^6 + 
(1-4\hat{\omega}^2)^3\cos(\omega T)\big)\,, \\
c_{-2} &= -\frac{128\eta_1\eta_2}{\pi(1-4\hat{\omega}^2)^3} \hat{\omega}
\sin(\omega T)
\,, \\
c_{-3/2} &= -\frac{\eta_2}{\sqrt{\pi}\hat{\omega}_-^3}
\left(\cos(BT \hat{\omega}_-^2/2) + \sin(BT \hat{\omega}_-^2/2)\right)
\\
& \qquad
 -\frac{\eta_2}{\sqrt{\pi}\hat{\omega}_+^3}
\left(\cos(BT \hat{\omega}_+^2/2) + \sin(BT \hat{\omega}_+^2/2)\right)
\,, \nonumber \\
c_{-1} &= \frac{4\eta_1^2}{\pi(1-4\hat{\omega}^2)^2}
   \big(1+4\hat{\omega}^2+(1-4\hat{\omega}^2)\cos(\omega T)\big) \,, \\
c_{-1/2} &= -\frac{\eta_1}{\sqrt{\pi}\hat{\omega}_-}
\left(\sin(BT \hat{\omega}_-^2/2) - \cos(BT \hat{\omega}_-^2/2)\right)
  \label{eq:Gmag1b}
\\
& \qquad
 -\frac{\eta_1}{\sqrt{\pi}\hat{\omega}_+}
\left(\sin(BT \hat{\omega}_+^2/2) - \sin(BT \hat{\omega}_+^2/2)\right)
\,. \nonumber
\end{align}
Here and below, $\hat{\omega} = \omega/B$, $\hat{\omega}_+ = \omega/B+1/2$, 
$\hat{\omega}_- = 1/2-\omega/B$,
and $0\leq \eta_1,\,\eta_2,\,\xi_1,\,\xi_2 < 1$.

If $\omega/B$ is in the neighborhood of $1/2$, namely, 
$|\omega/B-1/2| \ll \sqrt{{\pi}/{BT}}$, 
then
\begin{align}
\frac{B}{2\pi} |{\cal G}_\pm(\omega)|^2 &= \frac{c_{-3}}{(BT)^3} 
+ \frac{c_{-3/2}}{(BT)^{3/2}} +  \frac{c_{-1}}{BT}   
+ \frac{c_{-1/2}}{(BT)^{1/2}} + c_0 \,, 
  \label{eq:Gmag2}
\end{align}
where the coefficients are
\begin{align}
  \label{eq:Gmag2a}
c_{-3} &= \frac{32\eta_2^2}{\pi (1+2\hat{\omega})^6}\,, \\
c_{-3/2} &= -\frac{\eta_2}{6\sqrt{\pi}\hat{\omega}_+^3}
\left( 3(1+2\xi_1\sqrt{{BT}/{\pi}}\hat{\omega}_-)\cos(BT\hat{\omega}_+^2/2)
\right. \\
& \qquad\qquad\qquad\quad \left.
+ (3+\pi\xi_2(\sqrt{{BT}/{\pi}}\hat{\omega}_-)^3)\sin(BT\hat{\omega}_+^2/2) 
\right) \,, \nonumber \\
c_{-1} &= \frac{32\eta_1^2}{\pi(1+2\hat{\omega})^2} \,, \\
c_{-1/2} &= \frac{\eta_1}{6\sqrt{\pi}\hat{\omega}_+}
\left( 3(1+2\xi_1\sqrt{{BT}/{2}}\hat{\omega}_-)\sin(BT\hat{\omega}_+^2/2)
\right. \\
& \qquad\qquad\qquad\quad \left.
- (3+\pi\xi_2(\sqrt{{BT}/{2}}\hat{\omega}_-)^3)\cos(BT\hat{\omega}_+^2/2) 
\right) \,, \nonumber \\
c_{0} &= \frac{1}{4} 
 + \frac{\xi_1\sqrt{BT}\hat{\omega}_-}{2\sqrt{\pi}}
(1+\xi_1\sqrt{{BT}/{\pi}}\hat{\omega}_-) 
 + \frac{\pi\xi_2}{72}(\sqrt{{BT}/{\pi}}\hat{\omega}_-)^3
(6+\pi\xi_2(\sqrt{{BT}/{\pi}}\hat{\omega}_-)^3)
\,. 
  \label{eq:Gmag2b}
\end{align}
Here, the observation following Eq.~(\ref{eq:Sseries}) has been used.

In the region $\omega/B \geq 1/2$, if  
$\omega/B-1/2 \gg \sqrt{{\pi}/{BT}}$, 
then
\begin{align}
\frac{B}{2\pi} |{\cal G}_\pm(\omega)|^2 &= \frac{c_{-3}}{(BT)^3} + 
\frac{c_{-2}}{(BT)^2} + \frac{c_{-1}}{BT}\,, 
  \label{eq:Gmag3}
\end{align}
where
\begin{align}
  \label{eq:Gmag3a}
c_{-3} &= \frac{64 \eta_2^2}{\pi(1-4\hat{\omega}^2)^6}
\big(1+60\hat{\omega}^2 + 240\hat{\omega}^4 + 64\hat{\omega}^6 
+ (1-4\hat{\omega}^2)^3\cos(\omega T)\big)\,, \\
c_{-2} &= -\frac{128\eta_1\eta_2}{\pi(1-4\hat{\omega}^2)^3} \hat{\omega}
\sin(\omega T)
\,, \\
c_{-1} &= \frac{4\eta_1^2}{\pi(1-4\hat{\omega}^2)^2} \big(1+4\hat{\omega}^2
+(1-4\hat{\omega}^2)\cos(\omega T)\big) \,.
  \label{eq:Gmag4}
\end{align}

Eqs.~(\ref{eq:Gmag1})--(\ref{eq:Gmag4}) show that $|{\cal G}_\pm(\omega)|$ 
converges to the low-pass filter\\ $\sqrt{2\pi}\text{rect}(\omega/B)/\sqrt{B}$ 
as $T \to \infty$, with rigorously bounded corrections scaling as 
$1/\sqrt{T}$ and $1/\omega$.
In more detail, Eq.~(\ref{eq:Gmag1}) shows that at low frequencies
the magnitude is constant, up to corrections given explicitly in
Eqs.~(\ref{eq:Gmag1a})--(\ref{eq:Gmag1b}).
Likewise, the behaviour in the transition region is given by
Eqs.~(\ref{eq:Gmag2})--(\ref{eq:Gmag2b}).
Finally, Eq.~(\ref{eq:Gmag3}) shows that in the tail the magnitude
is zero, up to corrections given explicitly in 
Eqs.~(\ref{eq:Gmag3a})--(\ref{eq:Gmag4}).

\section{Example X Gate}
\label{xgate}

Section \ref{Subsec:Gate_Universality} describes methods for
implementing $X$ rotations, $Z$ rotations, and controlled-phase gates
through the variation of the configuration of the potential wells. 
In this appendix, we give a concrete example showing how one can implement 
an $X$ rotation by $\pi$ (that is, a Pauli $X$ gate) by varying the barrier 
height of a double-well potential.

Specifically, we use the potential
\begin{equation}
V(x) = \frac{V_1}{\cosh^2(x)} + \frac{V_2}{1+g\cosh^2(x)}
     + \frac{V_3}{(1+g\cosh^2(x))^2} \,,
\end{equation}
where
\begin{equation}
V_1 = \frac{g(g+2)}{4(1+g)^2} \,, \quad
V_2 = -4b^2(g+2) \,, \quad
V_3 = 4b(b+1)(1+g).
\end{equation}
If $b > 0$ and $b > \frac{g}{2(1+g)} + 1$, then this is 
\emph{quasi-exactly solvable} \cite{Chen}, a term meaning that
a subset of the spectrum is exactly solvable. 
The ground and first excited energies are then (in units where 
$\hbar = 2m = 1$)
\begin{align}
E_1 &= -\frac{(2+g-4b(1+g))^2}{4(1+g)^2} \,, \\
E_2 &= -\frac{(2+3g-4b(1+g))^2}{4(1+g)^2}.
\end{align}

If we lower the barrier and raise it back to its original height with a
time dependence that is in a Gevrey class of some finite order, then
we can invoke the adiabatic theorem of \cite{N93}, which guarantees
exponential convergence to perfect adiabaticity as a function of the
slowness of the variation (\S~\ref{Subsec:Gate_Universality}). 
In this case, up to exponentially small
errors, the dynamics induced by the variation is described by the $2
\times 2$ adiabatic-frame Hamiltonian $H_A$ given in \eq{HA}. The
off-diagonal matrix elements of $H_A$ are precisely zero here 
because the ground state is an even function of $x$,
the first excited state is an odd function of $x$, and $\frac{dH}{dt}$
has $x \to -x$ symmetry. Thus the unitary transformation induced is
\begin{eqnarray}
U & = & \exp \left( \int dt \left[ \begin{array}{cc} E_1(t) & 0 \\ 0 &
      E_2(t) \end{array} \right] \right) \nonumber \\
& = & e^{i \alpha}
\left[ \begin{array}{cc} 1 & 0 \\ 0 & e^{i \phi} \end{array} \right]
\,,
\end{eqnarray}
where
\begin{eqnarray}
\alpha & = & \int_0^\tau dt E_1(t)\,, \\
\phi & = & \int_0^\tau dt ( E_2(t) - E_1(t))\,.
\end{eqnarray}
Thus, up to an irrelevant global phase $\alpha$, $U$ is a $Z$
gate if $\phi = \pi$.

Concretely, we can achieve this as follows. Let $B(s)$ be the
bump function
\begin{equation}
\label{bump}
B(s) = \left\{ \begin{array}{cl} \exp \left( - \frac{1}{s(1-s)}
    \right), & 0 < s < 1, \\
0, & \textrm{otherwise}. \end{array} \right.
\end{equation}
This is in the Gevrey class of order $2$. One can 
vary the barrier height in time by setting
\begin{equation}
b(t/\tau) = 1 + \beta B(t/\tau)
\end{equation}
for some choices of $\beta$ and $\tau$. For example, one numerically
finds that by choosing $g = 0.01$, $\beta = 50$, and $\tau = 96.1602$,
one obtains $\phi = \pi$. 
Thus, this choice of parameters achieves a $Z$ gate in the eigenbasis
of the double well, which is an $X$ gate, if we interpret occupation
of the left (right) well as logical zero (one). 

\section{Example Z Gate}
\label{zgate}

The $Z$ gate can be analyzed in the limit in which the double-well
potential becomes two separate wells. Consider the convenient exactly 
solvable example in which  each well is 
a special case of the hyperbolic 
Poschl-Teller potential \cite{PT33} (and the Rosen-Morse potential 
\cite{RM32}), namely, 
\begin{equation}
V(x) = - \alpha^2 \frac{\lambda (\lambda - 1)}{\cosh^2 (\alpha x)}
\,,
\end{equation}
in units where $\hbar = m = 1$.
For $\lambda > 1$ this is an attractive potential well, and its ground-state
energy is 
\begin{equation}
E_0 = - \alpha^2 (\lambda-1)^2 \,.
\end{equation}
If we temporarily increase the well depth $\alpha^2$ for the
logical-one well, then this induces a Pauli Z rotation $e^{i Z \theta}$,
for some angle $\theta$. In particular, $\theta = \pi$ corresponds to
a standard $Z$ gate.

Concretely, we can vary $\alpha^2$ according to
\begin{equation}
\alpha^2(t) = \alpha_0^2 (1+\beta B(t/\tau)) \,,
\end{equation}
where $B$ is the bump function given in \eq{bump}, $\alpha_0^2$ is the
initial well depth, and $\beta$ is a parameter we choose. As we make
the process slower by increasing $\tau$, the diabatic error amplitudes
vanish as $\exp \left[ - \tau^{1/3} \right]$ 
(\S~\ref{Subsec:Gate_Universality}). 
To achieve a fixed target phase $\theta$ as we increase $\tau$, we must 
correspondingly decrease $\beta$. Specifically, a brief calculation yields
\begin{equation}
\beta = - \frac{\theta}{(\lambda-1)^2 \tau \alpha_0^2 \eta} \,,
\end{equation}
where
\begin{equation}
\eta = \int_0^1 ds \exp \left[ -\frac{1}{s (1-s)} \right] \simeq
7.0299 \times 10^{-3}.
\end{equation}

In reality, the wells will not be infinitely separated. The
corrections to the above analysis, in which we have assumed the wells
to be perfectly isolated, are of the order of the inner product between the
ground states of the two wells. The Poschl-Teller potential well is
well localized: for $|x| \gg 1/\alpha$, $V(x) \simeq
0$. Consequently, in the outer region $|x| \gg 1/\alpha$, the
ground-state wavefunction decays as $\exp[-\sqrt{-2 m E_0} x]$ (where
we have now included explicit dependence on $m$ but kept $\hbar =
1$). Thus the above approximation becomes exponentially good as the
separation between wells is increased.

\bibliography{bqpqft}

\end{document}